\documentclass[prb,twocolumn,unsortedaddress,superscriptaddress,showpacs]{revtex4-2}

\usepackage{multirow}
\usepackage{color}
\usepackage{graphicx}
\usepackage{amsmath,amsfonts,amsthm} % Math packages
\usepackage{physics,siunitx} 
\usepackage{booktabs,array}
\usepackage[T1]{fontenc} % european language fonts

\graphicspath{{./images/}}

\newcolumntype{C}[1]{>{\centering\arraybackslash}p{#1}}

\begin{document}

\title{Counter-intuitive Ferroelectric Property And Non-negligible Orbital Magnetic Moment In Cr/Cu Based Perovskite Metal-Organic Frameworks}
\author{Kunihiro Yananose}
 \email{ykunihiro@snu.ac.kr}
\affiliation{Center for Theoretical Physics, Department of Physics and Astronomy, Seoul National University, Seoul 08826, Republic of Korea}
\author{Jaejun Yu}
\affiliation{Center for Theoretical Physics, Department of Physics and Astronomy, Seoul National University, Seoul 08826, Republic of Korea}
\date{\today}

\begin{abstract}
Metal-organic frameworks (MOFs) possess a hybrid nature, combining the inorganic properties from the metal ions and the organic properties from the molecular linkers.
Stroppa, \textit{et al.}, showed that the perovskite-type MOF [C(NH$_2$)$_3$]M[(HCOO)$_3$] (M = Cr, Cu) exhibits the magneto-electric coupled multiferroicity [Angew. Chem. Int. Ed. 50, 5847 (2011) and Adv. Mater. 25, 2284 (2013)]. Moreover, their ferroelectricity arises from the hybrid improper mechanism, which also explains the magneto-electric coupling.
In this work, we further examine the electric and magnetic properties of [C(NH$_2$)$_3$]M[(HCOO)$_3$]. We find that the hybrid mode composed of non-polar modes induces purely electronic polarization even without the polar mode. The polar mode compensates for the purely electronic polarization. It leads to a counter-intuitive argument that the inversion of the polar mode rather enhances the polarization. We provide microscopic origin and macroscopic analysis for this polarization property.
In addition, we find that the orbital magnetic moment is comparable to the spin contribution in the Cu-based MOF. Finally, we establish the model for the orbital magnetic moment based on the perturbation theory.
\end{abstract}

\date{\today}
\maketitle

\section{Introduction}
Metal-organic frameworks (MOFs) are crystals in which the metal ions are connected with each other by organic molecules. Choice of organic linker allows the variety in their structures.
One of its classes, the porous MOFs hold a large portion of cavities in them. Focusing on their tunable porosity, applications on gas storage, catalysis, etc. are widely studied~\cite{furukawa_chemistry_2013,suh_hydrogen_2012}.
On the other hand, dense MOFs hold much smaller cavities in comparison to the porous MOFs as ordinary crystals. Instead, metal ions can play rather a significant role, and the emergence of orderings is concerned in the dense MOFs~\cite{cheetham_theres_2007,wang_perovskite-like_2004,ye_ferroelectric_2006}. 
The combination of organic-inorganic features can induce both magnetism and ferroelectricity simultaneously, \textit{i.e.,} multiferroicity. 
In some multiferroic materials, the ferroelectric order appears by the coupling to the magnetic order~\cite{van_aken_origin_2004,cheong_multiferroics_2007,malashevich_first_2008}. In such cases, belonging to one category of the `improper' ferroelectricity, control of the magnetic property by the electric field is expected, and vice versa.  
Thus, both of the magnetic and electric ferroic orders, their coupling, and the role of the structural deformation are important interests of dense MOFs~\cite{stroppa_electric_2011, picozzi_advances_2012, stroppa_hybrid_2013, di_sante_tuning_2013, ghosh_strain_2015, tian_high-temperature_2015, gomez-aguirre_room-temperature_2015, jain_switchable_2016, ptak_experimental_2016, gomez-aguirre_coexistence_2016, fan_electric-magneto-optical_2017}.

\begin{figure}[t]
\centering
\includegraphics[width=0.5\textwidth]{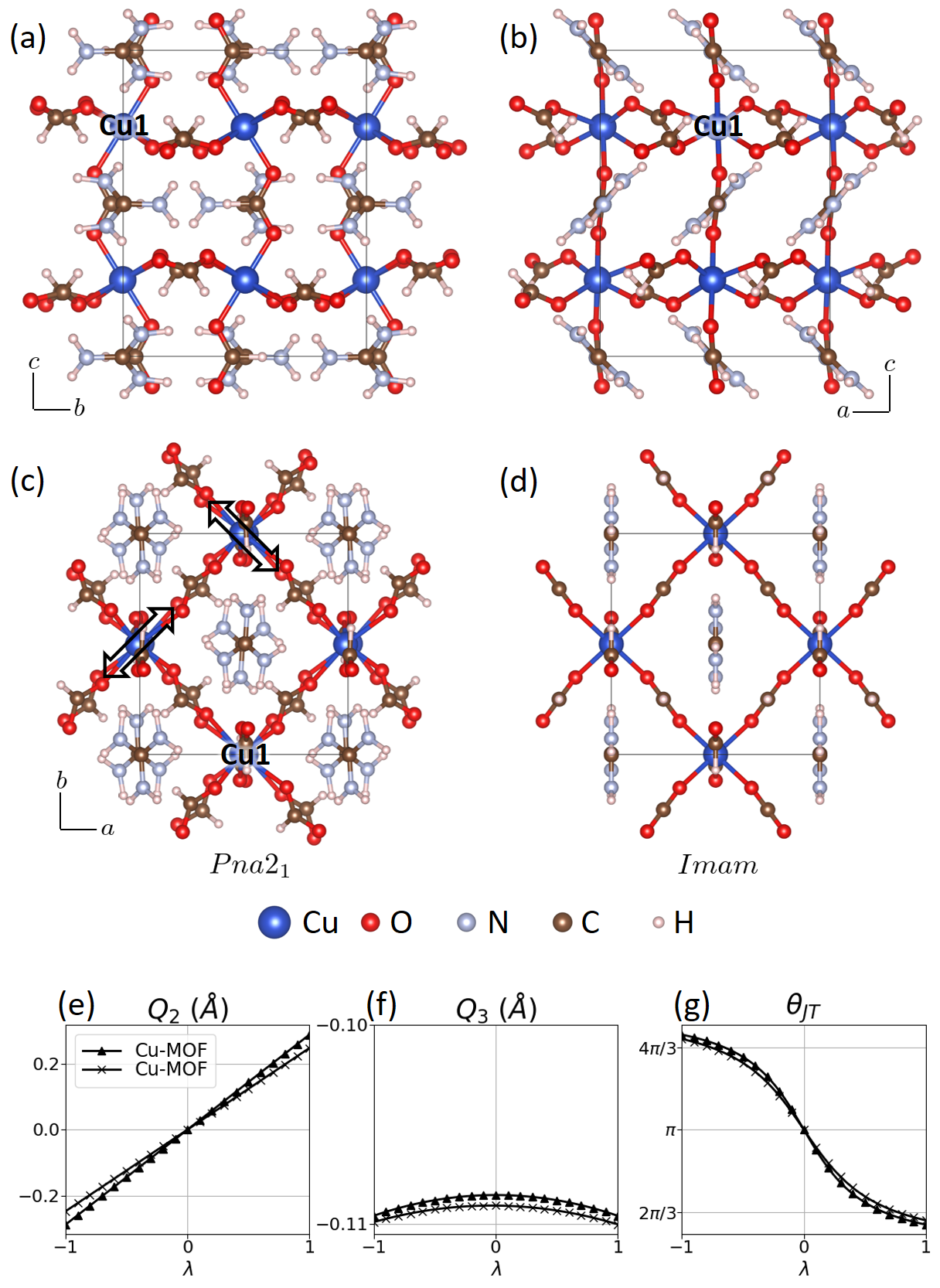}
\caption{
(a-c) $Pna2_1$ structure Cu-MOF ($\lambda=1$). The reference Cu1 is labeled. In (c), elongated directions are drawn by double-arrows. (d) $Imam$ structure Cu-MOF.
JT modes (e) $Q_2$ and (f) $Q_3$ of the reference Cu1 and Cr1 and their (g) JT phase are shown with respect to $\lambda$. 
}
\label{fig:system}
\end{figure}

Among the dense MOFs, [C(NH$_2$)$_3$]M[(HCOO)$_3$] (M = Cr, Mn, Fe, Co, Ni, Cu, and Zn) series have the perovskite-type ABX$_3$ structure~\cite{hu_metal-organic_2009, stroppa_hybrid_2013}. They consist of the guanidinium (Gua) ion (C(NH$_2$)$_3$)$^+$ for the A site, the $3d$ transition metal ions (M$^{2+}$) for the B site, and formate HCOO$^-$ ion for the X sites as shown in Fig.~\ref{fig:system} (a)-(c) for M = Cu case. These materials show the magnetic ordering by M$^{2+}$ ions. Especially, for the case in which the transition metal ion M is Jahn-Teller (JT) active ion Cr$^{2+}$ ($d^4$) or Cu$^{2+}$ ($d^9$), the hybrid improper ferroelectricity (HIFE) is theoretically predicted~\cite{stroppa_hybrid_2013}. In this paper, we will denote them as Cr- or Cu-MOF as following the previous studies. In the HIFE mechanism, the ferroelectric order parameter is not the primary order parameter. Instead, ferroelectric order appears by the trilinear coupling with two other primary order parameters~\cite{benedek_hybrid_2011}.
The electric polarization $P$ dependent part of the free energy is written as $F(P) = \alpha P^2 + \gamma Q_{X_a}Q_{X_b}P$, where $\alpha>0$. The second term is the trilinear coupling term including non-polar mode amplitudes $Q_{X_a}$ and $Q_{X_b}$, where $X_a$ and $X_b$ represent corresponding irreducible representations (irreps).
Then the spontaneous polarization appears as $P = -\gamma Q_{X_a}Q_{X_b}/2\alpha$. 
In addition, weak ferromagnetism (WFM) which arises by the canting of antiferromagnetically ordered spins is also reported by both the experiment and theory~\cite{stroppa_hybrid_2013,stroppa_electric_2011,tian_high-temperature_2015, hu_metal-organic_2009}. Therefore, Cr- and Cu-MOF are considered as the multiferroic materials carrying an electromagnetic coupling.

In this study, besides reproducing the known results from earlier studies~\cite{stroppa_electric_2011,stroppa_hybrid_2013}, we improve the arguments on both the electric and magnetic properties of the Cr-/Cu-MOF.
For the briefness, we will often refer to the electric polarization moment or its density as polarization in the rest of this paper, whereas we will not refer to the magnetic moment as the polarization to avoid confusion.
we perform density functional theory (DFT) calculation to obtain the energy, polarization, and magnetization of the Cr-/Cu-MOF with respect to the structures given by various combinations of the distortion modes.

Electric polarization consists of the core and electronic contribution. The core contribution comes from the positive point charge of atomic nuclei. In practice, the polarization from the core electrons which form closed shell near the nucleus is also included in the core contribution. The polarization from the rest valence electrons is the electronic contribution.
In an intuitive sense, the electric polarization moment is nearly proportional to the polar distortion mode. But this is not always true.
For example, in TbMnO$_3$, inversion symmetry breaking by the non-collinear spin spiral can induce purely electronic polarization even if the atoms are fixed in a non-polar structure~\cite{malashevich_first_2008}. 
We found an unusual properties of the Cr-/Cu-MOF that the hybrid mode of two non-polar modes induces non-negligible purely electronic polarization arising from the structural asymmetry and that the polar mode actually compensates for it. It leads to a counter-intuitive result that the reversal of the polar mode does not invert the total polarization, but rather enhances it. 

In general, the magnetic moment arises from two different origins, spin and orbital.
Usually, the orbital magnetic moment is small in comparison with the spin magnetic moment. For a transition metal ion in a ligand-octahedral environment, the orbital magnetic moment is quenched when the $t_{2g}$ $d$-orbitals are fully- or half-filled. This is why it is often ignored.
The previous studies~\cite{stroppa_electric_2011,stroppa_hybrid_2013} also ignored the orbital magnetic moment.
Even if the orbital magnetic moment is quenched, spin-orbit coupling (SOC) can induce a small orbital magnetic moment. However, it may not be negligible because the spin contribution arises as WFM. Actually, it turns out that the orbital contribution is comparable to the spin contribution in the magnetic moment of Cu-MOF.
We also construct a model to explain the orbital magnetic moment in the Cr-/Cu-MOF by combining the second order perturbation theory to the SOC and the orbital ordering described by the JT effective hamiltonian. A perturbative approach to the SOC was adopted to show the magnetic single ion anisotropy (MSIA) in the previous work~\cite{stroppa_hybrid_2013}. We will show that this model well matches the DFT results.

\begin{figure}[t]
\centering
\includegraphics[width=0.5\textwidth]{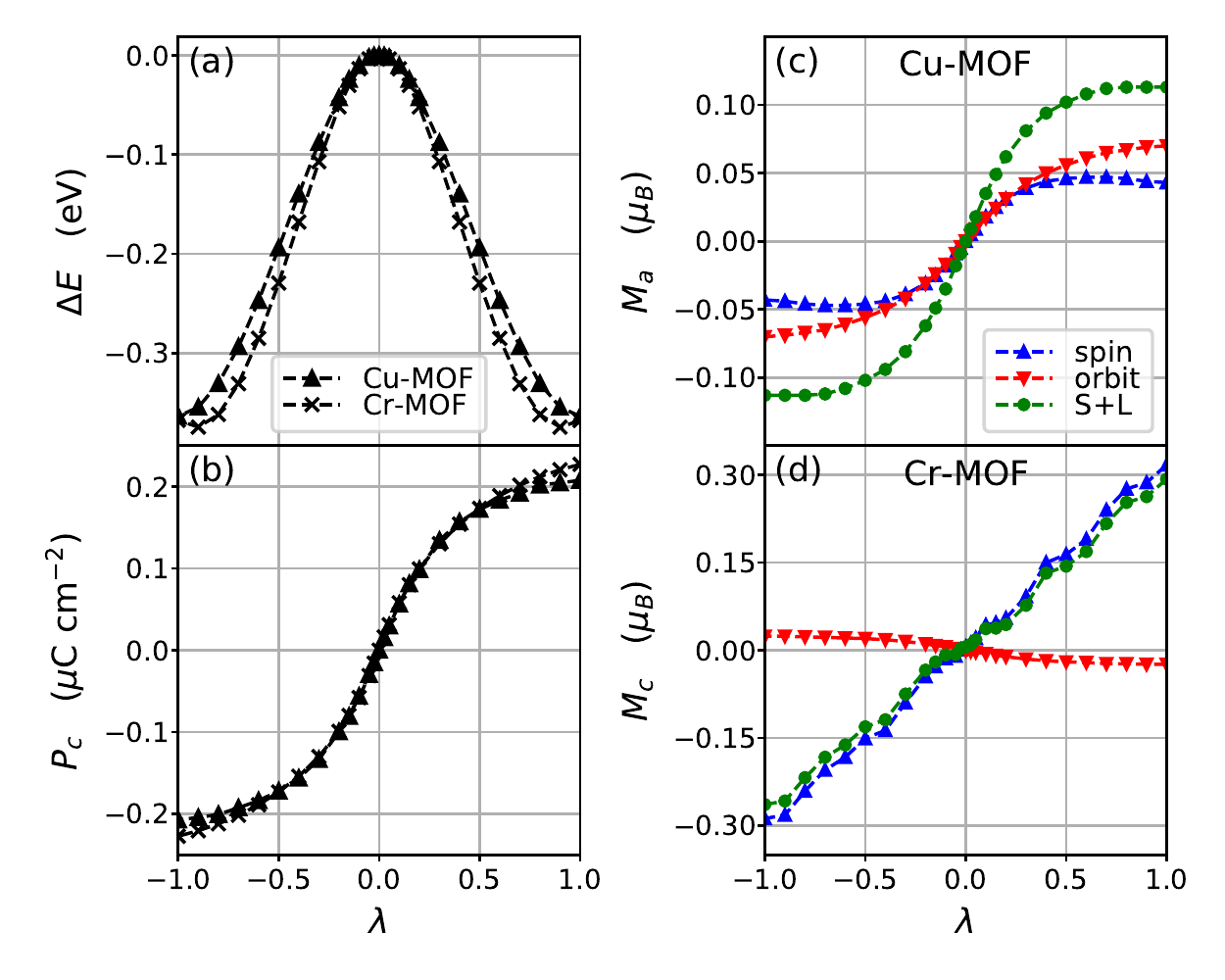}
\caption{
(a) Change of the energy and (b) electric polarization of the Cu-/Cr-MOF with respect to the structure parameter $\lambda$.
Spin, orbital, and total magnetic moment of (c) Cu-MOF and (d) Cr-MOF with respect to $\lambda$ of $Pnan$-path.
}
\label{fig:Pnan_path}
\end{figure}

\section{Methods}
We use the Vienna Ab initio Simulation Package (VASP)~\cite{kresse_efficient_1996} for the first-principles DFT calculation. To include SOC, we perform a non-collinear spin DFT calculation. Generalized gradient approximation by Perdew-Burke-Ernzerhof (GGA-PBE) for the exchange-correlation functional~\cite{perdew_generalized_1996} and the projector augmented wave pseudo-potential~\cite{kresse_ultrasoft_1999} are adopted. The plane wave energy cut-off is chosen to be 500 eV. $4\times4\times4$ regular $k$-space grid is used. To obtain the electric polarization moment in the periodic crystal, we used the Berry phase method~\cite{king-smith_theory_1993}. 
For the lattice constants, experimental values $a=8.5212$ \AA, $b=9.0321$ \AA, and $c=11.3497$ \AA\  from Ref.~\cite{hu_metal-organic_2009} are used.

For a ferroelectric structure, there exists a corresponding paraelectric virtual structure of higher symmetry. Such structure, referred to as pseudo symmetric structure, can be obtained by the group-theoretic method implemented in PSEUDO of the Bilbao Crystallography server~\cite{capillas_new_2011}.
Moreover, atomic displacements from the higher symmetry structure to the lower symmetry structure can be decomposed into orthogonal modes labeled after the irreps by using AMPLIMODE of Bilbao server~\cite{orobengoa_amplimodes_2009}.

\section{Structure, Magnetism, and Electric Polarization of C\lowercase{r}/C\lowercase{u}-MOF}
\subsection{$Pnan$ - path}
In this section, we will summarize the structure, magnetic, and electric properties of Cr-/Cu-MOF revealed by previous studies~\cite{stroppa_electric_2011,stroppa_hybrid_2013}, and then show our new findings on them. 
Cr$^{2+}$ and Cu$^{2+}$ ions are JT active. So the O$_6$ octahedron in which the ions are placed is significantly distorted. In the Cr-/Cu-MOF, JT distortions occur in alternating directions so that the elongated axis is perpendicular to the neighboring MO$_6$ octahedra as shown in Fig.~\ref{fig:system} (c). This antiferro-distortive structure induces the orbital ordering, \textit{i.e.}, a cooperative JT effect determines the orbital structure. The Goodenough-Kanamori-Anderson rule~\cite{goodenough_magnetism_1976,khomskii_transition_2014} predicts the ferromagnetic interaction between the in-plane neighboring ions and antiferromagnetic interaction between the out of plane neighboring ions. It results in the A-type antiferromagnetism (AFM-A). We assume that the major spin axis of AFM alignment is crystallographic $c$-axis for Cu-MOF and $a$-axis for Cr-MOF as following the previous studies.

The space group symmetry of JT distorted Cr-/Cu-MOF is the $Pna2_1$ (No. 33), which is a non-centrosymmetric group hosting a polarization. Corresponding pseudo symmetry group for Cr-/Cu-MOF in which the JT distortion is suppressed is the centrosymmetric $Pnan$ space group (No. 52, $Pnna$ in standard settings). It implies that the system gets a polarization by the deformation from the $Pnan$ to $Pna2_1$. The deformation can be expressed with the linear interpolation parameter $\lambda$. $\lambda=0$ means the $Pnan$ structure and $\lambda=1$ means the original $Pna2_1$ structure. 
Let's denote the atomic positions at $\lambda=0$ as $\mathbf{r}_{Pnan}$, and the displacement vectors from $\lambda=0$ structure to $\lambda=1$ structure as $\mathbf{u}$. Then the atomic positions of the interpolated structure are written as $\mathbf{r}_1(\lambda) = \mathbf{r}_{Pnan} + \lambda\mathbf{u}$. This displacement is labeled by single mode $\Gamma_4^-$. We will denote these interpolated structures as $Pnan$-path. 
The previous studies show the electric polarization appears monotonously to the $\lambda$. In this procedure, deformation of the Gua ions by hydrogen bond with the oxygen in formate ions induces the polarization.
As shown in Fig.~\ref{fig:Pnan_path} (a) and (b), we reproduced the change of the energy and the electric polarization of the Cr-/Cu-MOF with respect to the $\lambda$.

\begin{figure*}[t]
\centering
\includegraphics[width=0.8\textwidth]{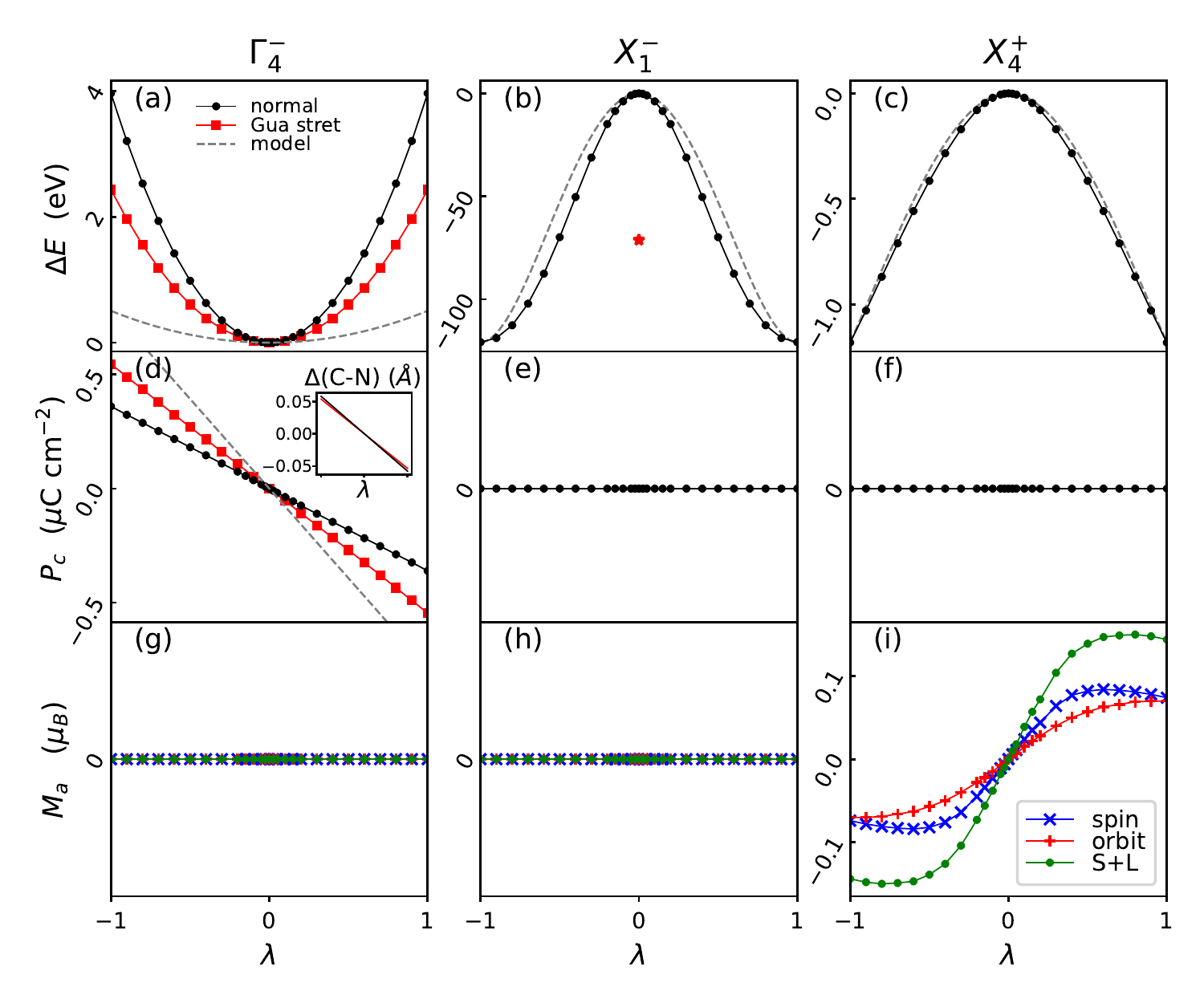}
\caption{
(a-c) Change of energy, (d-f) electric polarization, and (g-i) magnetic moments when only one of distortion mode among $\Gamma^-_4$, $X^-_1$ and $X^+_4$ exists. Each of the columns is the quantities changing with respect to the distortion mode labeled by the irrep on top of the column.
In (a-d), quantities obtained from the free energy model are shown in the grey dashed lines. 
In (a) and (d), red lines show the quantities calculated in the structure with the stretched Gua ion bonds. In (b), the red star indicates the relative energy of the structure with the stretched Gua ion bonds.
Inset in (d) is the C-N bond length difference of the Gua ion $\Delta$(C-N) $ = l^{\text{C-N}}_{\text{lower}}-l^{\text{C-N}}_{\text{upper}}$.
}
\label{fig:decomposed_1}
\end{figure*}

The experiments for the Cu-MOF observes the WFM moment, the remaining magnetic moment of canted spins from the AFM alignment~\cite{hu_metal-organic_2009}. The well-known mechanisms for the spin canting are Dzyaloshinskii-Moriya interaction (DMI)~\cite{dzyaloshinsky_thermodynamic_1958,moriya_anisotropic_1960} and the MSIA. Previous studies for Cr- and Cu-MOF shows that the DMI mechanism is discarded by the symmetry analysis. Instead, they show the MSIA justified by the second order perturbation theory treatment for the SOC induces the spin anisotropy. They show the switching of the magnetic moment along with the switching of $\lambda$ from 1 to -1. Moreover, the magnetic moment and the electric polarization shows nearly linear to each other implying that Cr- and Cu-MOF are electro-magnetic coupled multiferroics.

In this study, we examine the orbital magnetic moment which was overlooked in the previous studies. We found that in Cu-MOF the orbital magnetic moment is comparable to the spin magnetic moment and has the same direction with it. In the large $|\lambda|$ range, the orbital contribution is larger than the spin contribution.
On the other hand, for the Cr-MOF, the orbital moment is much smaller in comparison with the spin moment and has the opposite direction with it. Fig.~\ref{fig:Pnan_path} (c) and (d) show the calculated spin and orbital magnetic moment of Cr-/Cu-MOF.

\begin{figure}[t]
\centering
\includegraphics[width=0.5\textwidth]{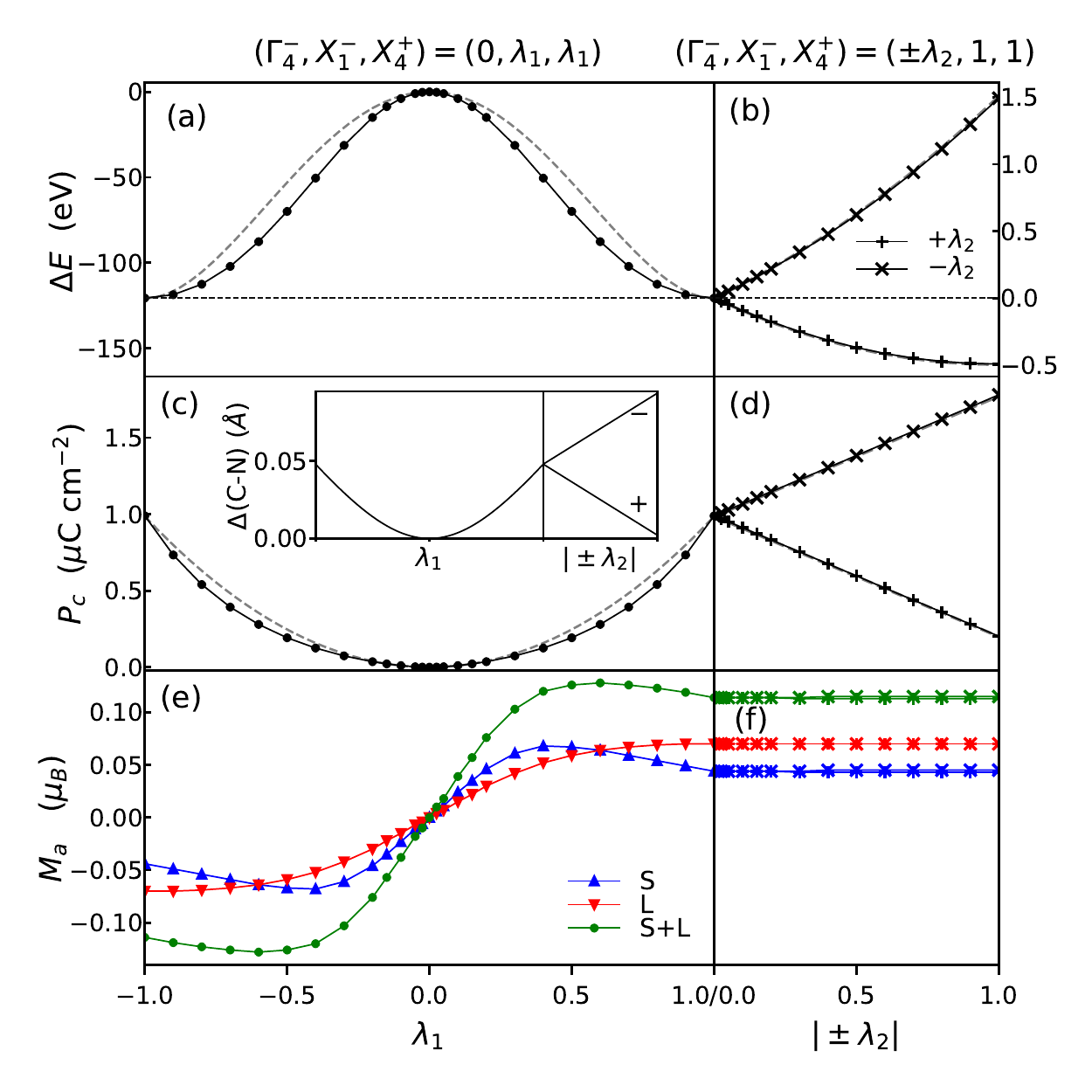}
\caption{
(a,b) Change of energy, (c,d) electric polarization, and (e,f) magnetic moments along the parameter path $(\lambda_{\Gamma^-_4},\lambda_{X^-_1},\lambda_{X^+_4}) = (0, \lambda_1, \lambda_1)$ and $(\lambda_{\Gamma^-_4},\lambda_{X^-_1},\lambda_{X^+_4}) = (\pm\lambda_2, 1, 1)$.
Be aware of the difference in the axis scale between (a) and (b).
Inset in (c) is the C-N bond length difference of the Gua ion $\Delta$(C-N).
Gray lines which show the free energy model derived values are perfectly overlapped with the DFT value in (b) and (d).  
}
\label{fig:decomposed_2}
\end{figure}

\subsection{$Imam$ - path}
Polar distortion in Cr-/Cu-MOF appears by coupling with the JT distortion as described by the hybrid improper ferroelectricity mechanism.
It can be seen by the deformation from a much higher symmetric structure, $Imam$ space group (No. 74, $Imma$). In this structure, all Gua ions are aligned parallel to each other in addition to the $Pnan$ structure. The displacements from $Imam$ structure to $Pna2_1$ structure can be decomposed into 3 orthogonal modes labeled after the irreducible representations $\Gamma^-_4$, $X^-_1$, and $X^+_4$.
$\Gamma^-_4$ mode is a polar mode. If only the $\Gamma^-_4$ mode is present, the structure has a $Ima2$ space group (No. 46).
$X^-_1$ mode corresponds to the rotation of Gua ions, resulting in the $Pnan$ space group.
$X^+_4$ mode mainly corresponds to the distortion of MO$_6$ octahedra (JT distortion) and includes a small distortion of the Gua ion, resulting in $Pnam$ space group (No. 62, $Pnma$).
The structure by the superposition of these 3 modes is, of course, $Pna2_1$. Interestingly, however, the combination of $X^-_1$ and $X^+_4$ modes without $\Gamma^-_4$ mode is already $Pna2_1$ structure, even though neither of $X^-_1$ and $X^+_4$ modes is a polar mode. We denote it as $X^-_1 \oplus X^+_4$ hybrid mode.
Similarly with the previous subsection, the structure can be described in terms of 3 linear interpolation parameters corresponding to each modes from the $Imam$ structure, $\mathbf{r}_2(\lambda_{\Gamma^-_4},\lambda_{X^-_1},\lambda_{X^+_4}) = \mathbf{r}_{Imam} + \lambda_{\Gamma^-_4}\mathbf{u}_{\Gamma^-_4} + \lambda_{X^-_1}\mathbf{u}_{X^-_1} + \lambda_{X^+_4}\mathbf{u}_{X^+_4}$, where the $\mathbf{r}_{Imam}$ is the atomic positions of $Imam$ structure, and $\mathbf{u}_\gamma$ and $\lambda_\gamma$ are distortion mode and its interpolation parameter corresponding to the irrep $\gamma$, respectively. We will denote these interpolated structures as $Imam$-path.

In Fig.~\ref{fig:decomposed_1}, the change of the energy, polarization, and magnetic moments of Cu-MOF with respect to each of the distortion modes are shown. The cases in which only single mode exists are considered.
The first row of Fig.~\ref{fig:decomposed_1} shows changes in energy. $X^-_1$ and $X^+_4$ modes are unstable modes which means that the energy of the system decreases by these modes. But $\Gamma^-_4$ mode is a stable mode that raises the energy.
The second row shows the electric polarization. Polar mode $\Gamma^-_4$ induces the electric polarization which is linear to it as shown in Fig.~\ref{fig:decomposed_1} (d). However, its sign is opposite to the electric polarization of the final $Pna2_1$ structure. Non-polar mode $X^-_1$ and $X^+_4$ do not induce the polarization.
The last row shows the magnetic moments. Only the $X^+_4$ mode representing the JT distortion induces non-zero total magnetic moment (Fig.~\ref{fig:decomposed_1} (i)). It means that the canting of the spin is coupled with the JT phase.

Because the symmetry operation based method searching for the pseudo symmetry structure is weak at capturing the rotation of molecules, the bond lengths of Gua are significantly shortened. To examine the realistic situations, we also calculate the quantities with the structure in which the bond lengths of Gua are stretched to reasonable values for some cases.
For $\Gamma^-_4$ mode, stretched Gua bond length results in the smaller energy change and the larger polarization as shown by the red lines in Fig.~\ref{fig:decomposed_1} (a) and (d).
For $X^-_1$ mode, the energy change is reduced almost half by the Gua bond stretching.

As a following step, we examined the structure path from $(\lambda_{\Gamma^-_4},\lambda_{X^-_1},\lambda_{X^+_4}) = (0,0,0)$ to $(0,1,1)$, and from $(0,1,1)$ to $(\pm 1,1,1)$. They correspond to the displacement paths from $Imam$ structure to original $Pna2_1$ structure and to its polar mode inverted structure $(-1,1,1)$, which are decomposed into the $X^-_1 \oplus X^+_4$ mode and $\Gamma^-_4$ mode.
The results are shown in Fig.~\ref{fig:decomposed_2} with the parameter $\lambda_1$ for $X^-_1 \oplus X^+_4$ mode and $\pm\lambda_2$ for $\pm\Gamma^-_4$ mode.
Obviously, the energy of the system decreases by the combination of two unstable modes, $X^-_1 \oplus X^+_4$ as shown in Fig.~\ref{fig:decomposed_2} (a). The change in energy is symmetric between the positive and negative sides of $\lambda_1$. In the presence of the $X^-_1 \oplus X^+_4$ hybrid modes, $\Gamma^-_4$ mode reduces the energy which was the stable mode in the absence of it (Fig.~\ref{fig:decomposed_2} (b)). It defines the HIFE, \textit{i.e.}, polar $\Gamma^-_4$ mode appears via the coupling with the $X^-_1 \oplus X^+_4$ mode~\cite{stroppa_hybrid_2013}.
On the other hand, inverted $\Gamma^-_4$ mode increases the energy. it implies that $(\lambda_{\Gamma^-_4},\lambda_{X^-_1},\lambda_{X^+_4}) = (1,1,1)$ and $(-1,1,1)$ structures are not energetically equivalent, thus they are not related with each other by any symmetry operation. Actually, $(\lambda_{\Gamma^-_4},\lambda_{X^-_1},\lambda_{X^+_4}) = (\lambda,1,\lambda)$ structure in $Imam$-path corresponds to $\mathbf{r}_1(\lambda)$ structure of $Pnan$-path, \textit{i.e.}, in addition to the polar mode $\Gamma^-_4$, $X^+_4$ mode also has to be inverted to obtain the $\mathbf{r}_1(\lambda=-1)$ structure of $Pnan$-path. This is an important feature of HIFE mechanism.

Magnetic moment by the $X^-_1 \oplus X^+_4$ mode shown in Fig.~\ref{fig:decomposed_2} (e) exhibit a similar tendency with the $X^+_4$ mode only case. The spin magnetic moment is slightly reduced in the presence of $X^-_1$ mode. $\Gamma^-_4$ mode has no effect on the magnetic moment (Fig.~\ref{fig:decomposed_2} (f)).

Because the $X^-_1 \oplus X^+_4$ hybrid mode results in polar space group $Pna2_1$, electric polarization appears even without the polar $\Gamma^-_4$ mode as shown in Fig.~\ref{fig:decomposed_2} (c). Moreover, there is no core contribution to the polarization, \textit{i.e.}, the polarization is purely electronic. Because both of the $X^-_1$ and  $X^+_4$ modes are non-polar, the changes in the core contribution are all canceled.
The polarization is also symmetric between the positive and negative sides of $\lambda_1$. It means that if we switch both of the $X^-_1$ and $X^+_4$ modes, polarization is not switched. It is a consistent result with the Ref.~\cite{benedek_hybrid_2011}.
As shown in Fig.~\ref{fig:decomposed_2} (d), additional electric polarization by $\pm\Gamma^-_4$ mode is linear to $|\lambda_2|$ as in the $\Gamma^-_4$ mode only case (Fig.~\ref{fig:decomposed_1} (d)). Now, the core contribution is present. The sign of the change in polarization is also consistent with it. $\Gamma^-_4$ mode induces the moment opposite to the final moment, \textit{i.e.}, the main origin of the electric polarization is $X^-_1 \oplus X^+_4$ hybrid mode and polar $\Gamma^-_4$ mode reduces it but does not invert the sign of the total moment.
It results in an interesting result, the inversion of the polar $\Gamma^-_4$ mode does not invert the electric polarization, but rather enhance it.
We confirm the same properties in the Cr-MOF.

\begin{figure}[t]
\centering
\includegraphics[width=0.5\textwidth]{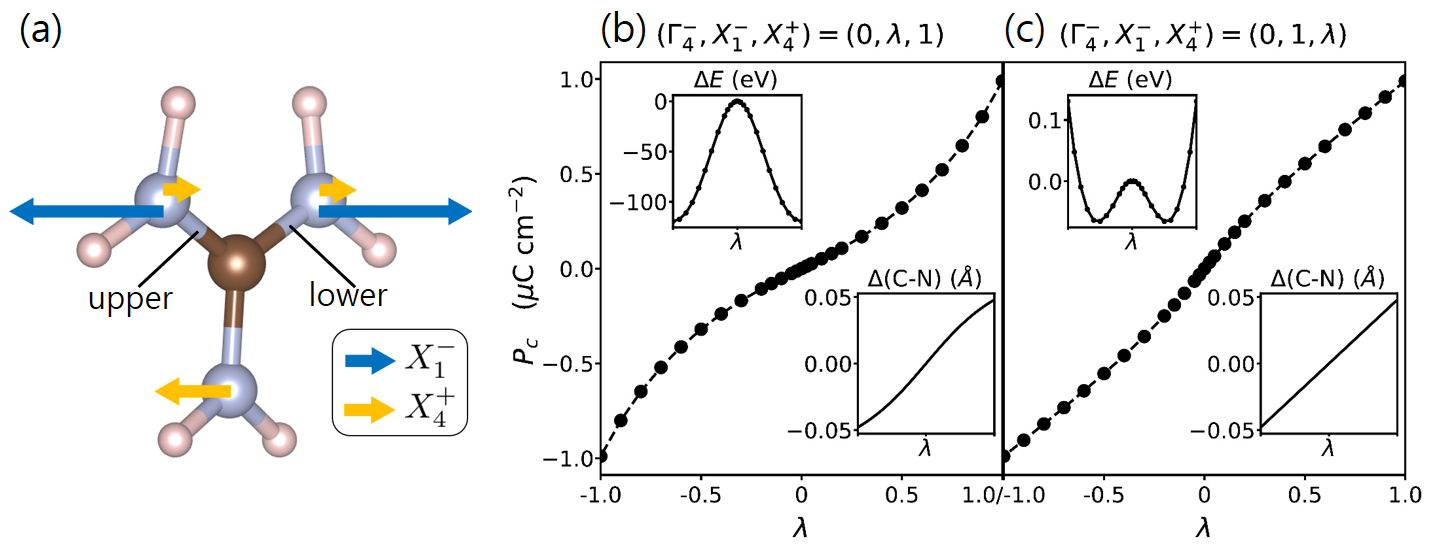}
\caption{
(a) Schematic picture of the displacements of N atoms relative to the C atom in Gua by the $X^-_1$ and $X^+_4$ modes. Polarization without $\Gamma^-_4$ mode in the (b) $(\lambda_{\Gamma^-_4},\lambda_{X^-_1},\lambda_{X^+_4}) = (0,\lambda,1)$ path and (c) $(\lambda_{\Gamma^-_4},\lambda_{X^-_1},\lambda_{X^+_4}) = (0,1,\lambda)$ path. In (b) and (c), upper insets are energy change and lower insets are the C-N bond length difference of the Gua ion $\Delta$(C-N).
}
\label{fig:Gua_HIFE}
\end{figure}

Tian, \textit{et al.}, suggested that the A-site Gua ions are an important factor of the polarization~\cite{tian_high-temperature_2015}. According to them, in terms of Lewis formalism, a Gua$^{+}$ ion has one double bond out of three carbon-nitrogen (C-N) bonds as a resonant state. A localized positive charge is considered to be at the N connected by the double bond. In this picture, the shorter bond takes the higher probability that the double bond is placed at it. As a result, the C-N bond length difference of the Gua ion induces the polarization.
The microscopic origin of the purely electronic polarization is also attributed to this mechanism. The bond length difference between the Gua's lower and upper C-N bonds with respect to $c$-axis, $\Delta$(C-N) $ = l^{\text{C-N}}_{\text{lower}}-l^{\text{C-N}}_{\text{upper}}$, is consistent with the polarization. In the inset of Fig.~\ref{fig:decomposed_2} (c), $\Delta$(C-N) is shown. Two bonds are equivalent in the $Imam$ phase so that the system is non-polar. In the presence of the distortion, the $\Delta$(C-N) is nearly quadratic and symmetric to $\lambda_1$ and linear to $-\lambda_2$. It is exactly the characteristic of the corresponding polarization. Moreover, the same is also true for the $\Gamma^-_4$ mode only case, in which the $\Delta$(C-N) is shown in the inset of Fig.~\ref{fig:decomposed_1} (d). 

On the other hand, in either $X^-_1$ or $X^+_4$ mode only cases, $\Delta$(C-N) remains zero. Each of these modes contains the alternating rotation of Gua ions, rotation around $b$-axis by $X^-_1$ mode and around $c$-axis by $X^+_4$. Note that they are not ideal rotation so the bond lengths change by them. By single non-polar mode, N atoms move symmetrically with respect to the C atom, so that the $\Delta$(C-N) unchanged. However, if two modes coexist, the combined displacement of N is no longer symmetric to the C atom. As a result, $\Delta$(C-N) becomes finite. This is depicted in Fig.~\ref{fig:Gua_HIFE} (a) in which the displacements of N atoms relative to C atom by $X^-_1$ and $X^+_4$ modes are shown. In this way, the combination of the two non-polar modes can give rise to a polar space group.
It also implies that the prediction of the polarization by the naive point charge assumption may not be applied for the purely electronic polarization because it is perpendicular to the atomic displacements in Gua.

We further analyze the purely electronic polarization by the $X^-_1 \oplus X^+_4$ hybrid mode.
By the arguments on the hybrid improper FE, switching of either $X^-_1$ or $X^+_4$ mode, but not both, inverts the polarization. Ref.~\cite{benedek_hybrid_2011} exhibits the HIFE by showing the polar mode is frozen in the presence of the hybrid mode. But they also suggested the possibility of the ferroelectric state by the hybrid mode only.
We calculate the polarization without the polar $\Gamma^-_4$ mode in the $(\lambda_{\Gamma^-_4},\lambda_{X^-_1},\lambda_{X^+_4}) = (0,\lambda,1)$ path and $(\lambda_{\Gamma^-_4},\lambda_{X^-_1},\lambda_{X^+_4}) = (0,1,\lambda)$ path as shown in Fig.~\ref{fig:Gua_HIFE} (b) and (c). In both cases, the polarization is inverted by the switching of one of the non-polar mode. Again, $\Delta$(C-N) is consistent with the polarization in these cases. On the other hand, the fact that switching both of the non-polar modes does not inverts the polarization was already shown in Fig.~\ref{fig:decomposed_2} (c).
These observations imply that the purely electronic polarization without the polar $\Gamma^-_4$ mode is attributed to the HIFE mechanism.
The same argument is also applied to the $\Gamma^-_4$ mode. The structures energetically equivalent to the $(\lambda_{\Gamma^-_4},\lambda_{X^-_1},\lambda_{X^+_4}) = (1,1,1)$ are $(\lambda_{\Gamma^-_4},\lambda_{X^-_1},\lambda_{X^+_4}) = (-1,-1,1)$,$(-1,1,-1)$, and $(1,-1,-1)$. It is equivalent to the HIFE argument, \textit{i.e.}, $\Gamma^-_4$ mode is inverted by switching either $X^-_1$ or $X^+_4$ mode, but not both.
Then we can say that two different physical quantities, the purely electronic polarization and the polar distortion $\Gamma^-_4$ mode, are simultaneously coupled with two non-polar modes $X^-_1$ or $X^+_4$.

The coupling between the polarization and the distortion mode can be represented by Landau theory~\cite{landau_statistical_2008}.
In terms of the usual HIFE mechanism, $P$ dependent part of the free energy is written as $F(P) = \alpha P^2 + \gamma Q_{X_1^-}Q_{X_4^+}P$. The switching rule of HIFE comes from the spontaneous polarization described as $P = -\gamma Q_{X_1^-}Q_{X_4^+}/2\alpha$.
Now we construct the free energy for Cr-/Cu-MOF in which the polarization $P$ and the polar $\Gamma^-_4$ mode are described separately. Because the polarization and the polar distortion mode respect the same symmetry, both of them can have the terms in the same form in the free energy. In addition, a linear coupling of them $PQ_{\Gamma_4^-}$ can be included, which actually has the same symmetry with $P^2$ term.
The $P$ and $Q_{\Gamma_4^-}$ dependent part of the free energy is
\begin{equation}
\label{eq:landau1}
\begin{split}
F(P,Q_{\Gamma_4^-}) &= \alpha P^2 + \alpha'Q_{\Gamma_4^-}^2 + \beta PQ_{\Gamma_4^-}\\
&+ \gamma Q_{X_1^-}Q_{X_4^+}P + \gamma' Q_{X_1^-}Q_{X_4^+}Q_{\Gamma_4^-} 
\end{split}
\end{equation}
where $\alpha>0$ and $\alpha'>0$ are assumed.
Then the spontaneous polarization is
\begin{equation}
\label{eq:P1}
P^* = -\frac{\beta}{2\alpha} Q_{\Gamma_4^-} -\frac{\gamma}{2\alpha} Q_{X_1^-}Q_{X_4^+} .
\end{equation}
If $\beta>0$, $\gamma<0$, and $\beta/\alpha<-\gamma/\alpha$, the first-principles results are well explained. Without $\Gamma_4^-$ mode, $\lambda_1$ dependency of $P^*$ is $P^*\sim Q_{X_1^-}Q_{X_4^+}\sim\lambda_1^2$. In addition to it, $P^*$ is linear to $Q_{\Gamma_4^-}\sim \lambda_{(2)}$.
If we replace the $P$ in Eq.~(\ref{eq:landau1}) with the Eq.~(\ref{eq:P1}), 
\begin{equation}
\label{eq:landau_G4}
\begin{split}
F(Q_{\Gamma_4^-}) 
&=-\frac{\gamma^2}{4\alpha}(Q_{X_1^-}Q_{X_4^+})^2 \\
&+(\gamma'-\frac{\beta\gamma}{2\alpha})Q_{X_1^-}Q_{X_4^+}Q_{\Gamma_4^-}
+(\alpha'-\frac{\beta^2}{4\alpha})Q_{\Gamma_4^-}^2 .
\end{split}
\end{equation}
When $(\gamma'-\beta\gamma/2\alpha)<0$ and $(\alpha'-\beta^2/4\alpha)>0$, freezing of the $\Gamma_4^-$ mode shown in Fig.~\ref{fig:decomposed_2} (b) is also reproduced. 
Determination of the parameters is described in Appendix~\ref{sec:app_param}.

In addition to the polarization dependent terms, the free energy includes the elastic energies due to the non-polar modes,
\begin{equation}
\label{eq:non-polar}
\begin{split}
F_{\text{non-polar}}(Q_{X_1^-}, Q_{X_4^+}) &= \eta Q_{X_1^-}^2 + \eta' Q_{X_4^+}^2 \\
&+ \lambda Q_{X_1^-}^4 +\lambda' Q_{X_4^+}^4 +\xi Q_{X_1^-}^2Q_{X_4^+}^2.
\end{split}
\end{equation}
These terms are necessary to determine the switching field strength (See Appendix~\ref{sec:app_nonpol}).
Finally, the total energies and polarization derived from the model with fitted parameters are shown together in Fig.~\ref{fig:decomposed_1} and \ref{fig:decomposed_2}.

We can construct a detailed argument on the HIFE and magneto-electric coupling in the Cu-MOF.
The polarization can be decomposed into two parts, the hybrid mode part by the $X^-_1 \oplus X^+_4$ mode and the polar mode part by the $\Gamma^-_4$ mode which is coupled with $X^-_1 \oplus X^+_4$ mode.
From the $Imam$ structure, $X^-_1 \oplus X^+_4$ mode first appears and induces both of the magnetic and electric polarization moment. Next, the $\Gamma^-_4$ mode appears to partially compensate for the polarization and further stabilize the energy. This is why reversing the polar mode rather enhances the polarization.  
Magneto-electric coupling which is experimentally confirmed~\cite{tian_high-temperature_2015} is rather clearly explained by $X^+_4$ mode. Because, both the electric and magnetic moments vary with $X^+_4$ mode, whereas $\Gamma^-_4$ mode does not change the magnetic moment.
Furthermore, we will show that the orbital angular momentum is explicitly coupled with the JT distortion represented by $X^+_4$ mode.

\section{Model for the orbital magnetic moment in C\lowercase{r}-/C\lowercase{u}-MOF}
\subsection{Spin-Orbit Coupling Hamiltonian and Jahn-Teller Distortion}
In the previous study for Cr-MOF, the second order perturbation theory for the SOC was adopted to explain the spin canting~\cite{stroppa_hybrid_2013}.
To explain the orbital magnetic moment in Cr-/Cu-MOF, we establish the model in which the perturbation method is combined with the JT effective hamiltonian within single ion description. 
The perturbation approach for the orbital angular momentum and MSIA is basically Bruno theory~\cite{bruno_tight-binding_1989,blanco-rey_validity_2019}, but we ignore the k-space dispersion for simplicity. 

To express the SOC Hamiltonian $H_{\text{SOC}} = \zeta\vec{\mathbf{S}}\cdot\vec{\mathbf{L}}$, let's denote the local coordinate unit vectors for spin operator as $(\mathbf{x}',\mathbf{y}',\mathbf{z}')$, and that for orbital angular momentum as $(\mathbf{x},\mathbf{y},\mathbf{z})$. We rotate the primed coordinate with respect to the unprimed coordinate according to the Euler angle rule. Then the primed coordinate unit vectors are $\mathbf{x}' = \cos\theta\cos\phi\mathbf{x} + \cos\theta\sin\phi\mathbf{y} - \sin\theta\mathbf{z}$, $\mathbf{y}' = -\sin\phi\mathbf{x} + \cos\phi\mathbf{y}$, and $\mathbf{z}' = \sin\theta\cos\phi\mathbf{x} + \sin\theta\sin\phi\mathbf{y}+ \cos\theta\mathbf{z}$.
SOC Hamiltonian is written as
\begin{equation}
\label{eq:H_SOC}
\begin{split}
H_{\text{SOC}} &= \zeta\vec{\mathbf{S}}\cdot\vec{\mathbf{L}}\\
&= \frac{\zeta}{2}
\begin{pmatrix}
\sin\theta\cos\phi & \cos\theta\cos\phi + i\sin\phi \\
\cos\theta\cos\phi - i\sin\phi & - \sin\theta\cos\phi \\ 
\end{pmatrix}
L_{x}\\
&+ \frac{\zeta}{2}
\begin{pmatrix}
\sin\theta\sin\phi & \cos\theta\sin\phi - i\cos\phi \\
\cos\theta\sin\phi + i\cos\phi & - \sin\theta\sin\phi \\ 
\end{pmatrix}
L_{y}\\
&+ \frac{\zeta}{2}
\begin{pmatrix}
\cos\theta & -\sin\theta \\
-\sin\theta & -\cos\theta \\ 
\end{pmatrix}
L_{z}.\\
\end{split}
\end{equation}
The matrix representation for the orbital angular momentum operator is determined by the quantum mechanical relations for the angular momentum states $L_{z}\ket{l,m} = m\ket{l,m}$ and $L_{\pm}\ket{l,m} = \sqrt{(l\mp m)(l\pm m +1)}\ket{l,m\pm 1}$, where we adopt the atomic units in which $\hbar=1$.
We will consider only $d$-orbitals here~\cite{takayama_magnetic_1976}.

For the transition metal ion in the O$_6$ octahedron cage, $d$-orbitals are energetically separated into lower energy $t_{2g}$ orbitals $(d_{yz},d_{zx},d_{xy})$ and higher energy $e_g$ orbitals $(d_{x^2-y^2},d_{z^2})$ by the crystal field splitting. If the spin configuration allows the degrees of freedom between the degenerated orbitals, the system tends to lower its energy by deforming the O$_6$ cage and splitting the degeneracy of $e_g$ orbitals, \textit{i.e.}, JT effect~\cite{jahn_stability_1937,khomskii_transition_2014}.
Deformation of the octahedron is represented by two distortion modes $Q_2$ and $Q_3$,
\begin{equation}
\begin{split}
Q_2 &= \frac{1}{\sqrt{2}}(l_{x}-l_{y})\\
Q_3 &= \frac{1}{\sqrt{6}}(2l_{z}-l_{x}-l_{y})
\end{split}
\end{equation}
where the $l_i$ means the distance from the center of the octahedron to the oxygen on the $i$-axis. Then the JT distorted structure is expressed with the JT phase $\theta_{\text{JT}}$ as follows.
\begin{equation}
\begin{split}
&\ket{\theta_{\text{JT}}} = 
\cos\theta_{\text{JT}}\ket{Q_3} +
\sin\theta_{\text{JT}}\ket{Q_2} \\
&\tan\theta_{\text{JT}} = \frac{Q_2}{Q_3}.
\end{split}
\end{equation}
The JT effective Hamiltonian taking the $e_g$ orbitals as a basis is given by
\begin{equation}
\begin{split}
&H_{\text{JT}} = 
\gamma\begin{pmatrix}
q_1 & q_2 \\
q_2 & -q_1 \\
\end{pmatrix}
+\frac{1}{2}Cq^2\mathbf{I}_2 \\
\end{split}
\end{equation}
where $q_1=q\cos(\theta_{\text{JT}}/2)$, $q_2=q\sin(\theta_{\text{JT}}/2)$, and $\mathbf{I}_2$ is $2\times2$ identity matrix~\cite{stroppa_analogies_2016}.
The energy eigenvalues are $E_{\pm} = \pm \gamma q + \frac{1}{2}Cq^2$ and eigenstates are
\begin{equation}
\begin{split}
\ket{d_{-}(\theta_{\text{JT}})}&=
-\sin\left(\theta_{\text{JT}}/2\right)\ket{d_{x^2-y^2}}
+\cos\left(\theta_{\text{JT}}/2\right)\ket{d_{z^2}} \\
\ket{d_{+}(\theta_{\text{JT}})}&=
\cos\left(\theta_{\text{JT}}/2\right)\ket{d_{x^2-y^2}}
+\sin\left(\theta_{\text{JT}}/2\right)\ket{d_{z^2}}.\\
\end{split}
\end{equation}
It represents the orbital-JT phase locking.
These eigenstates can be considered to be unitary rotated $e_g$ orbitals according to the JT effective Hamiltonian.
Then, the unitary matrix is
\begin{equation}
\begin{split}
U_0 =
\begin{pmatrix}
-\sin\left(\frac{\theta_{\text{JT}}}{2}\right) 
& \cos\left(\frac{\theta_{\text{JT}}}{2}\right) \\
\cos\left(\frac{\theta_{\text{JT}}}{2}\right) 
& \sin\left(\frac{\theta_{\text{JT}}}{2}\right) \\
\end{pmatrix}
\end{split}
\end{equation}
The unitary matrix for the whole $d$-orbitals is
\begin{equation}
\begin{split}
U =
\begin{pmatrix}
\mathbf{I}_3 & \mathbf{0} \\
\mathbf{0} & U_0 \\
\end{pmatrix}
\end{split}
\end{equation}
Consequently, the newly defined orbital angular momentum operator matrices considering JT effect can be obtained by unitary rotation with this matrix, $(L_i)^{\text{new}} = U^{\dagger}(L_i)^{\text{old}}U$.

\subsection{Perturbation Theory and Orbital Angular Momentum}
The perturbation theory is applied to obtain the orbital angular momentum by the SOC. The `JT transformed' $d$-orbitals given by the unitary transform of the previous subsection are taken as the unperturbed basis ${\ket{d^0_{n\sigma}}}$ where $n=\{yz,zx,xy,-,+\}$ and $\sigma= \uparrow\text{ or }\downarrow$ spins.
The first order corrected $d$-orbitals are
\begin{equation}
\ket{d_{\alpha}} = \ket{d^0_{\alpha}} + 
\sum_{\beta\neq\alpha}\frac{\mel{d^0_{\beta}}{H_{SOC}}{d^0_{\alpha}}}
{E^0_{\alpha}-E^0_{\beta}}\ket{d^0_{\beta}} 
\end{equation}
where the $\alpha$ and $\beta$ are combined indices of orbital species and spin. 
We can obtain the orbital angular momentum of transition metal ion in the JT distorted O$_6$ cage by calculating $L_i = \sum_{\alpha\in\text{occ}}\mel{d_{\alpha}}{(L_{i}^{d})^{\text{new}}}{d_{\alpha}}$ up to first order in $\zeta$.
In the rest of this paper, the superscripts `$d$' and `new' of the orbital angular momentum operator are omitted.
For the spin-up high spin configuration of the Cr$^{2+}$ ion, occupied $d$-orbitals are $\{d_{yz\uparrow},d_{zx\uparrow},d_{xy\uparrow},d_{-\uparrow}\}$, and for the Cu$^{2+}$ ion $\{d_{yz\uparrow},d_{zx\uparrow},d_{xy\uparrow},d_{-\uparrow},d_{+\uparrow},d_{yz\downarrow},d_{zx\downarrow},d_{xy\downarrow},d_{-\downarrow}\}$.

The orbital angular momentum expectation value for a perturbed $d$-orbital $d_{n\uparrow}$ is
\begin{equation}
\label{eq:Li_dn}
\begin{split}
\mel{d_{n\uparrow}}{L_i}{d_{n\uparrow}} &= \mel{d^0_{n\uparrow}}{L_i}{d^0_{n\uparrow}} \\
&+ \sum_{m\neq n}\Big[\tfrac{\mel{d^0_{m\uparrow}}{H_{SOC}}{d^0_{n\uparrow}}}{E^0_{n\uparrow}-E^0_{m\uparrow}}
\mel{d^0_{n\uparrow}}{L_i}{d^0_{m\uparrow}}+c.c.\Big]\\
&+ \sum_{\text{all }m}\Big[\tfrac{\mel{d^0_{m\downarrow}}{H_{SOC}}{d^0_{n\uparrow}}}{E^0_{n\uparrow}-E^0_{m\downarrow}}
\mel{d^0_{n\uparrow}}{L_i}{d^0_{m\downarrow}}+c.c.\Big]\\
&+ O(\zeta^2).\\ 
\end{split}
\end{equation}
Because $\mel{d^0_{n\uparrow}}{L_i}{d^0_{n\uparrow}} = \mel{d^0_{n\uparrow}}{L_i}{d^0_{m\downarrow}} = 0$, only the second term remains up to the first order in $\zeta$. 
For each $x,y,z$ component, summation over occupied orbitals in $d^4$ configuration are
\begin{equation}
\label{eq:Lxyz}
\begin{split}
(L_x)_{d^4}
= &- \left(\frac{\zeta}{E^0_{+}-E^0_{yz}}\right)
\sin\theta\cos\phi \\
&\times(\cos(\tfrac{\theta_{\text{JT}}}{2}) 
+\sqrt{3}\sin(\tfrac{\theta_{\text{JT}}}{2}))^2 \\ 
(L_y)_{d^4}
= &- \left(\frac{\zeta}{E^0_{+}-E^0_{zx}}\right)
\sin\theta\sin\phi \\
&\times(\cos(\tfrac{\theta_{\text{JT}}}{2}) 
-\sqrt{3}\sin(\tfrac{\theta_{\text{JT}}}{2}))^2
\\
(L_z)_{d^4}
= &- \left(\frac{\zeta}{E^0_{+}-E^0_{xy}}\right)
\cos\theta
(2\cos(\tfrac{\theta_{\text{JT}}}{2}))^2 .
\\
\end{split}
\end{equation}
For $d^9$ configuration, $(L_i)_{d^9} = -(L_i)_{d^4}$.

\subsection{Model Analysis for Cr-/Cu-MOF}
Due to the negative sign of the electron charge, both the spin and orbital magnetic moment have opposite directions to corresponding angular momenta. If we replace the angular momentum with the magnetic moment for both of the spin and orbital, \textit{i.e.}, $\theta$ and $\phi$ indicate the direction of the spin magnetic moment, and $\mathbf{L}$ is read as the orbital magnetic moment in the Bohr magneton $\mu_\text{B}$ unit, Eq.~(\ref{eq:Lxyz}) are still valid for the magnetic moment. In the later part of this work, we will use Eq.~(\ref{eq:Lxyz}) in the magnetic moment sense.

To calculate the total orbital magnetic moment of the Cr-/Cu-MOF, we take one of 4 Cr/Cu ions in a unit cell of MOF, say Cr/Cu1, as a reference to describe the system. If we know the local MO$_6$ structure and the local moment of Cr/Cu ion at one site, those of other sites are automatically determined by the space group and magnetic group symmetry. For these systems with AFM-A order, two magnetic groups are allowed, $Pna'2_1'$ and $Pn'a'2_1$~\cite{stroppa_electric_2011}, where the prime means that the symmetry operation is accompanied by the time-reversal operation. In $Pna'2_1'$ ($Pn'a'2_1$), AFM spin axis is crystallographic $c$ ($a$)-axis and weak FM canting direction is $a$ ($c$) direction, that is corresponding to Cu-MOF (Cr-MOF) in our case. In the total magnetic moment, only $a$ ($c$) component remains non-zero and other components are canceled with the moments of other sites for $Pna'2_1'$ ($Pn'a'2_1$). The transformation rules of the magnetic moment and the corresponding Cr/Cu site numbers with respect to Cr/Cu1 by the symmetry operations of the magnetic group $Pna'2_1'$ and $Pn'a'2_1$ are listed in Table.~\ref{tab:table1} in terms of the crystallographic axes.

\begin{figure}[t]
\centering
\includegraphics[width=0.5\textwidth]{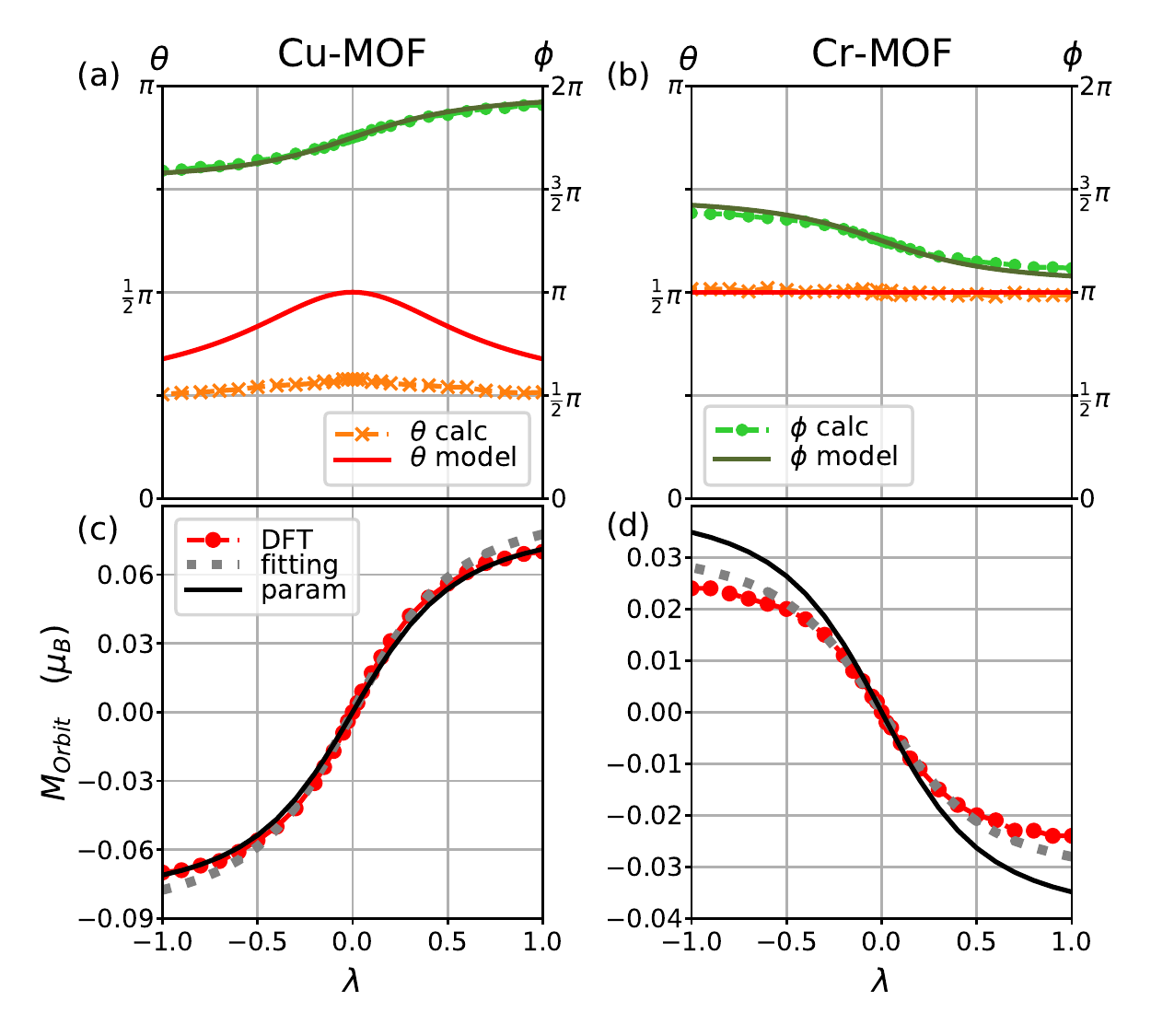}
\caption{
The direction of the orbital magnetic moment of (a) Cu1 of Cu-MOF and (b) Cr1 of Cr-MOF in their local coordinates obtained from DFT and the model. 
The total orbital magnetic moment of (c) Cu-MOF and (d) Cr-MOF obtained from DFT and model.
Model values in (c) and (d) are fitted to the DFT results (gray dotted line) and evaluated from reasonable physical parameters (black solid line).
}
\label{fig:orb_mom_model_1}
\end{figure}

\begin{table}
\begin{tabular}{C{0.06\textwidth}C{0.035\textwidth}C{0.14\textwidth}C{0.035\textwidth}C{0.14\textwidth}}
\hline
\multirow{2}{*}{Cr/Cu} & \multicolumn{2}{c}{$Pna'2_1'$} & \multicolumn{2}{c}{$Pn'a'2_1$}\\
\cmidrule{2-5}
 & op. & $\vec{L}$ & op. & $\vec{L}$ \\
\hline
\hline
 1 & $1$ & $(L_a,L_b,L_c)$      & $1$ & $(L_a,L_b,L_c)$ \\
 2 & $n$ & $(L_a,-L_b,-L_c)$    & $n'$ & $(-L_a,L_b,L_c)$ \\
 3 & $a'$ & $(L_a,-L_b,L_c)$    & $a'$ & $(L_a,-L_b,L_c)$ \\
 4 & $2'_1$ & $(L_a,L_b,-L_c)$  & $2_1$ & $(-L_a,-L_b,L_c)$ \\
\hline
\end{tabular}
\caption{
Labels of Cr/Cu ions, corresponding symmetry operations, and the transformation rules of the magnetic moment by them in the magnetic space group $Pna'2_1'$ and $Pn'a'2_1$. $1$ means the identity operation.}
\label{tab:table1}
\end{table}

Let us consider the O$_6$ octahedron of the reference Cr/Cu1 ion. The local coordinate of the orbital magnetic moment is aligned to its O-M bond directions. Our coordinate is determined by the following steps. First, put the octahedron in the way that the local coordinate ($x,y,z$) are aligned with crystallographic ($\hat{a},\hat{b},\hat{c}$) direction. Then, rotate the octahedron by $-\pi/4$ around the $\hat{c}$ axis, and by tilting angle $-\theta_{\text{t}}$ around the $\hat{a}$ axis consecutively. Then the local coordinate with respect to the $(\hat{a},\hat{b},\hat{c})$ is given by
\begin{equation}
\label{eq:coordinate_trans}
\begin{split}
\hat{x} &= \tfrac{1}{\sqrt{2}}\hat{a} - \tfrac{1}{\sqrt{2}}\cos\theta_{\text{t}}\hat{b} + \tfrac{1}{\sqrt{2}}\sin\theta_{\text{t}}\hat{c} \\
\hat{y} &= \tfrac{1}{\sqrt{2}}\hat{a} + \tfrac{1}{\sqrt{2}}\cos\theta_{\text{t}}\hat{b} -\tfrac{1}{\sqrt{2}}\sin\theta_{\text{t}}\hat{c} \\
\hat{z} &= \sin\theta_{\text{t}}\hat{b}+\cos\theta_{\text{t}}\hat{c} \\
\end{split}
\end{equation}
Now, we can express the direction of the local spin magnetic moment in terms of the Euler angle $\theta_{\text{spin}}$ and $\phi_{\text{spin}}$ with respect to the local coordinate of the orbital magnetic moment.
If we ignore a small spin canting, the direction of spin is exactly $c$-direction in $Pna'2_1'$ magnetic group. It correspond to $\theta_{\text{spin}}=\theta_{\text{t}}$ and $\phi_{\text{spin}}=-\tfrac{\pi}{4}$. For $Pn'a'2_1$, spin direction is $a$ and corresponding angles are $\theta_{\text{spin}}=\tfrac{\pi}{2}$ and $\phi_{\text{spin}}=\tfrac{\pi}{4}$.
The geometry of O$_6$ octahedra of Cr/Cu2$\sim$4 and their magnetic moments are derived from those of Cr/Cu1 by the magnetic symmetries.

For the $d^4$ configuration (Cr$^{2+}$) with $Pna'2_1'$ symmetry, the total moment is 4 times of the $a$-component of the moment of the reference Cr1.
From Eq.~(\ref{eq:Lxyz}) and Eq.~(\ref{eq:coordinate_trans}),
\begin{equation}
\begin{split}
L_{\text{total}} 
= & 4L_{a}
= 4(\tfrac{1}{\sqrt{2}}L_{x}+\tfrac{1}{\sqrt{2}}L_{y}) \\
=& - 2\left(\tfrac{\zeta}{E^0_{+}-E^0_{yz}}
-\tfrac{\zeta}{E^0_{+}-E^0_{zx}}\right)
\sin\theta_{\text{t}}
(2-\cos\theta_{\text{JT}})\\
& - 2\sqrt{3}\left(\tfrac{\zeta}{E^0_{+}-E^0_{yz}}
+\tfrac{\zeta}{E^0_{+}-E^0_{zx}}\right)
\sin\theta_{\text{t}}
\sin\theta_{\text{JT}}.\\
\end{split}
\end{equation}
Because the difference between $E^0_{yz}$ and $E^0_{zx}$ will be small, the first term will be small in comparison with the second term.
Moreover, if $\theta_{\text{JT}} = \pi$ which corresponds to $\lambda_{X^+_4} = 0$, $E^0_{yz} \approx E^0_{zx}$ because the Cu/Cr-O bond length along $x$- and $y$- local axes will be equivalent. As a result, $L_{\text{total}}$ vanishes when $\lambda_{X^+_4} = 0$. It is consistent with the first principles result.
By introducing assumption $E^0_{yz}=E^0_{zx}=E^0_{xy}\equiv E^0_{t_{2g}}$, the orbital magnetic moment can be simplified as
\begin{equation}
\label{eq:Ltot_simple}
\begin{split}
L_{\text{total}} 
&= - 4\sqrt{3}\left(\tfrac{\zeta}{E^0_{+}-E^0_{t_{2g}}}\right)
\sin\theta_{\text{t}}
\sin\theta_{\text{JT}}.\\
\end{split}
\end{equation}
Likewise, for the $d^4$ with $Pn'a'2_1$ symmetry, the total orbital magnetic moment is 4 times of the $c$-component of the moment of the reference Cr1.
\begin{equation}
\begin{split}
L_{\text{total}} 
= & 4L_{c}
= 4(\tfrac{1}{\sqrt{2}}\sin\theta_{\text{t}}L_{x}-\tfrac{1}{\sqrt{2}}\sin\theta_{\text{t}}L_{y}+\cos\theta_{\text{t}}L_{z}) \\
=& - 2\left(\tfrac{\zeta}{E^0_{+}-E^0_{yz}}
-\tfrac{\zeta}{E^0_{+}-E^0_{zx}}\right)
\sin\theta_{\text{t}}
(2-\cos\theta_{\text{JT}})\\
& - 2\sqrt{3}\left(\tfrac{\zeta}{E^0_{+}-E^0_{yz}}
+\tfrac{\zeta}{E^0_{+}-E^0_{zx}}\right)
\sin\theta_{\text{t}}
\sin\theta_{\text{JT}}.\\
\end{split}
\end{equation}
Interestingly, the same formula with the $Pna'2_1'$ case is obtained. Therefore, the same arguments are also valid and it results in the same simplified form of Eq.~(\ref{eq:Ltot_simple}).
For the $d^9$ configuration (Cu$^{2+}$), the sign of the orbital magnetic moment is inverted in both magnetic groups.

As a preliminary for the comparison between the DFT calculation results and the predictions from the model, we parametrize the JT phase of reference Cr/Cu1 as a function of $\lambda$ of the $Pnan$-path. In the Cu-MOF, as the $\lambda$ increases from 0 to 1, Q$_2$ changes linearly from 0 to 0.288 as shown in Fig.~\ref{fig:system} (e). Meanwhile, Q$_3$ changes very little, so that it can be considered as a constant (Fig.~\ref{fig:system} (f)). $\lambda=0$ and $\lambda=1$ correspond to $\theta_{\text{JT}}=\pi$ and $\theta_{\text{JT}}=1.934\approx0.616\pi$, respectively (Fig.~\ref{fig:system} (g)). Then, $\tan(\pi-\theta_{\text{JT}})$ is proportional to Q$_2$. Finally, the following parametrization can be obtained.
\begin{equation}
\frac{\tan(\pi-\theta_{\text{JT}})}{\tan(\pi-\theta_{\text{JT,}\lambda=1})} = \lambda
\end{equation}
where $\theta_{\text{JT,}\lambda=1}$ is the JT phase at $\lambda=1$.
For the simplicity, let's assume $\theta_{\text{JT,}\lambda=1}=2\pi/3$ which corresponds to $\ket{d^0_{-}}=-\ket{x^2}$ and $\ket{d^0_{+}} = -\ket{y^2-z^2}$ instead of $0.616\pi$. JT phase of the Cr-MOF can be represented by the same parametrization with Cu-MOF. 
Then, we can get the following expression for the JT phase in terms of the $\lambda$.
\begin{equation}
\theta_{\text{JT}} = \pi - \tan^{-1}(\sqrt{3}\lambda)
\end{equation}
Then, the simplified orbital magnetic moment Eq.~(\ref{eq:Ltot_simple}) can be written in terms of the $\lambda$,
\begin{equation}
\begin{split}
(L_{\text{total}})_{d^4/d^9} &= \mp 4\sqrt{3}\left(\tfrac{\zeta}{E^0_{+}-E^0_{t_{2g}}}\right)\sin\theta_{\text{t}}\sin\theta_{\text{JT}}\\
 &= \mp 4\sqrt{3}\left(\tfrac{\zeta}{E^0_{+}-E^0_{t_{2g}}}\right)\sin\theta_{\text{t}}\frac{\sqrt{3}\lambda}{\sqrt{3\lambda^2+1}}.
\end{split}
\end{equation}

Ignoring the JT phase dependency of $E^0_i$'s, we define the $\lambda$ independent part of this expression as $A$,
\begin{equation}
A = \mp 4\sqrt{3}\left(\tfrac{\zeta}{E^0_{+}-E^0_{t_{2g}}}\right)
\sin\theta_{\text{t}}.
\end{equation}
To check the validity of the model, we compare the orbital magnetic moment calculated from the DFT, the model with $A$ obtained from fitting to DFT results, and the model with $A$ obtained from reasonable physical parameters as shown in Fig.~\ref{fig:orb_mom_model_1} (c) and (d).
Fitted $A$ values are 0.090 for Cu-MOF and -0.032 for Cr-MOF.
For the parameters, we adopted SOC parameter $\zeta= 56.54$ meV for Cu and $\zeta= 27.52$ meV for Cr~\cite{griffith_theory_2009}, $\Delta E = E^0_{+}-E^0_{t_{2g}} = 2.5$ eV for both of the Cr-MOF and Cu-MOF, and tilting angle of MO$_6$ octahedron $\theta_{\text{t}} = 31.61^\circ$ for Cu-MOF and $\theta_{\text{t}}=31.88^\circ$ for Cr-MOF. The resultant $A$ values are 0.082 for Cu-MOF and -0.040 for Cr-MOF.
These are reasonably consistent with the DFT results.

In addition, let us consider the direction of the orbital magnetic moment of Cu1 and Cr1 expected from Eq.~(\ref{eq:Lxyz}) with the assumption $E^0_{yz}=E^0_{zx}=E^0_{xy}$. They are shown in Fig.~\ref{fig:orb_mom_model_1} (a) and (b) in their local spherical coordinates with the direction from the DFT for the comparison. Except for the deviation in the polar angle $\theta$ of Cu1, the model well predicts the orbital magnetic moment direction.

\section{Conclusion}
In this work, we highlighted unusual aspects of both the electric and magnetic properties of Cr- and Cu-MOFs [C(NH$_2$)$_3$]M[(HCOO)$_3$] and provided an improved understanding.
On the electronic property, the hybrid mode $X^-_1 \oplus X^+_4$ which is the combination of two non-polar modes induces purely electronic polarization even without the polar mode $\Gamma^-_4$. In the microscopic viewpoint, bond length asymmetry in the Gua ions induces purely electronic polarization. The polar mode $\Gamma^-_4$ compensates for the polarization stabilizing the energy. It results in an interesting property. Contrary to common-sense, if we invert the polar mode $\Gamma^-_4$, the polarization is rather enhanced. In the macroscopic viewpoint of Landau theory, these unusual electric properties can be described by the doubly hybrid improper mechanism in which the polarization and the polar mode order parameters are treated separately.
We expect that our approach can be applied to other materials exhibiting HIFE.  

On the magnetic property, we found that the orbital magnetic moment is comparable to the spin contribution in the Cu-MOF. Even though the orbital magnetic moment is quenched, SOC induces a finite orbital magnetic moment. To explain the orbital magnetic moment, we established the model in which the perturbative approach to the SOC is combined with the JT transformed orbital angular momentum operator. It must be generally applicable to JT active $d^4$ and $d^9$ configuration in the ligand-octahedron environment. Although the orbital magnetic moment is small, it can be comparable to the spin contribution in the WFM materials as in Cu-MOF.  

\section{Acknowledgement}
Authors thank Alessandro Stroppa for useful information and fruitful discussions.
This work was supported by Samsung Electronics Co., Ltd.

\begin{widetext}

\section{Appendix}

\subsection{Determination of The Parameters in Free Energy}
\label{sec:app_param}
In this appendix section, we determine the coefficients of the free energy of Cu-MOF, Eq.~(\ref{eq:landau1}).
To determine the values of the parameters in free energy, let us set a rule for the units. The free energy is measured in eV per unit cell. For simplicity, we use the DFT total energy values as free energy. Polarization is written as the polarization density in the $\mu$C/cm$^2$ unit. Distortion mode amplitudes are replaced with the dimensionless ratio to their values in equilibrium in $Pna2_1$ structure, \textit{i.e.}, $Q_X$ becomes equivalent to $\lambda_X$.
For the $\Gamma^-_4$ mode related parameters, two different data can be used to determine the same parameters, from $(\lambda_{\Gamma^-_4},\lambda_{X^-_1},\lambda_{X^+_4}) = (\lambda,0,0)$ path and from $(\lambda_{\Gamma^-_4},\lambda_{X^-_1},\lambda_{X^+_4}) = (\lambda,1,1)$ path. However, the resulting parameters from two data are incompatible. In such cases, we choose the data from the structure which is closer to the equilibrium $Pna2_1$ structure, $(\lambda,1,1)$.
Instead, this choice brings about relatively large error in $(\lambda,0,0)$ path shown in Fig.~\ref{fig:decomposed_1} (a) and (d).  

From the polarization values, we get
\begin{equation}
-\frac{\gamma}{2\alpha} = 0.99\ \mu C/\text{cm}^2
\end{equation}
with $(\lambda_{\Gamma^-_4},\lambda_{X^-_1},\lambda_{X^+_4}) = (0,\lambda,\lambda)$ path of Fig.~\ref{fig:decomposed_2} (c) and
\begin{equation}
-\frac{\beta}{2\alpha} = -0.79\ \mu C/\text{cm}^2
\end{equation}
with $(\lambda_{\Gamma^-_4},\lambda_{X^-_1},\lambda_{X^+_4}) = (\lambda,1,1)$ path of Fig.~\ref{fig:decomposed_2} (d).
From the energy change in $(\lambda_{\Gamma^-_4},\lambda_{X^-_1},\lambda_{X^+_4}) = (\lambda,1,1)$ path of Fig.~\ref{fig:decomposed_2} (b), we get
\begin{equation}
\begin{split}
&(\gamma'-\frac{\beta\gamma}{2\alpha})+(\alpha'-\frac{\beta^2}{4\alpha})
=-0.493470\ \text{eV}\\
&-(\gamma'-\frac{\beta\gamma}{2\alpha})+(\alpha'-\frac{\beta^2}{4\alpha})
=1.49337\ \text{eV}.
\end{split}
\end{equation}
However, we have four equations for five parameters. It is impossible to determine the parameters from the given data.

Instead, we estimate the $\alpha$ value from the separate calculations. The $\alpha$ represents the energy from the polarization and it is highly attributed to Gua ions. Therefore, we estimate $\alpha$ by applying external electric field $\mathcal{E}$ to isolated symmetric Gua$^+$ ions and calculating the energy and induced dipole moment $p$. In the calculation by using VASP, an external electric field is added by sawtooth type potential and the +1 oxidation number is realized by reducing one electron with NELECT option.
The energy of such system depending on $p$ and $\mathcal{E}$ is written as $E(p,\mathcal{E}) = ap^2 - p\mathcal{E}$.
By fitting to the calculated values shown in Fig.~\ref{fig:landau_alpha}, we obtain $a=1.175$ eV/(e\AA)$^2$.
Because Cr-/Cu-MOF has four Gua$^+$ ions in a unit cell, the relation between $a$ and $\alpha$ can be given as $\alpha (p/v)^2 = 4ap^2$  where $v = 873.522$~\AA$^3$ is the volume of unit cell. 
As a result, we get
\begin{equation}
\begin{split}
&\alpha = 1.40\ (\text{eV}/[\mu C/\text{cm}^2]^2)\\
&\alpha' = 1.37\ (\text{eV})\\
&\beta = 2.20\ (\text{eV}/[\mu C/\text{cm}^2])\\
&\gamma = -2.76\ (\text{eV}/[\mu C/\text{cm}^2])\\
&\gamma' = -3.17\ (\text{eV}) .
\end{split}
\end{equation}

\begin{figure}[t]
\centering
\includegraphics[width=0.6\textwidth]{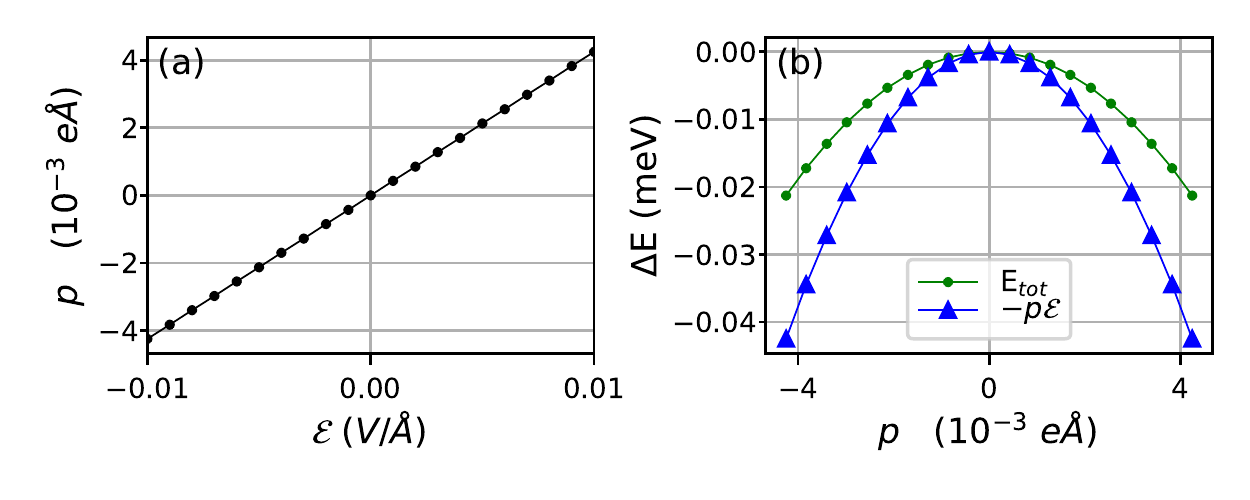}
\caption{
(a) Induced dipole moment in isolated Gua$^+$ ion by external electric field $\mathcal{E}$. (b) Total energy change and dipole-field interaction energy.
}
\label{fig:landau_alpha}
\end{figure}

\subsection{Estimation of The Ferroelectric Switching Field}
\label{sec:app_nonpol}
In the presence of the external field $h$, free energy for the second order phase transition as a function of order parameter $x$ can be written as $f(x) = \alpha x^2 + \beta x^4 -xh$, where $\alpha<0$ and $\beta>0$. When this free energy has minima $-y_0$ at $\pm x_0$, $\alpha = -2y_0/x_0^2$, $\beta= y_0/x_0^4$, and the switching field strength is $|h_c| = 4|\alpha|^{3/2}/(3\sqrt{6\beta}) = 8y_0/(3\sqrt{3}x_0)$.
The free energy of Cu-MOF including external electric field $\mathcal{E}$ is
\begin{equation}
\label{eq:landau_field}
F(P,Q_{\Gamma_4^-},\mathcal{E}') = \alpha P^2 + \alpha'Q_{\Gamma_4^-}^2+ \beta PQ_{\Gamma_4^-}
+ \gamma Q_{X_1^-}Q_{X_4^+}P + \gamma' Q_{X_1^-}Q_{X_4^+}Q_{\Gamma_4^-} -P\mathcal{E}',
\end{equation}
where $\mathcal{E}' = v\mathcal{E}$. The resultant polarization is
\begin{equation}
P^* = -\frac{\beta}{2\alpha} Q_{\Gamma_4^-} -\frac{\gamma}{2\alpha} Q_{X_1^-}Q_{X_4^+} +\frac{1}{2\alpha}\mathcal{E}' .
\end{equation}
When it is substituted to Eq.~(\ref{eq:landau_field}),
\begin{equation}
\label{eq:landau_field2}
\begin{split}
F(P^*,Q_{\Gamma_4^-},\mathcal{E}') 
&=-\frac{\gamma^2}{4\alpha}(Q_{X_1^-}Q_{X_4^+})^2
+(\gamma'-\frac{\beta\gamma}{2\alpha})Q_{X_1^-}Q_{X_4^+}Q_{\Gamma_4^-}
+(\alpha'-\frac{\beta^2}{4\alpha})Q_{\Gamma_4^-}^2 \\
&+\frac{\gamma}{2\alpha}Q_{X_1^-}Q_{X_4^+}\mathcal{E}' +\frac{\beta}{2\alpha}Q_{\Gamma_4^-}\mathcal{E}' -\frac{1}{4\alpha}\mathcal{E}'^2 .
\end{split}
\end{equation}
The $Q_{\Gamma_4^-}$ determined by the given $Q_{X_1^-}$ and $Q_{X_4^+}$ is,
\begin{equation}
\begin{split}
Q^*_{\Gamma_4^-} = -\frac{(\gamma'-\frac{\beta\gamma}{2\alpha})}{2(\alpha'-\frac{\beta^2}{4\alpha})}Q_{X_1^-}Q_{X_4^+}
-\frac{\beta}{4\alpha(\alpha'-\frac{\beta^2}{4\alpha})}\mathcal{E}'
\equiv
\frac{E_1}{2E_2}Q_{X_1^-}Q_{X_4^+} - \frac{P_2}{2E_2}\mathcal{E}',
\end{split}
\end{equation}
where $P_1 \equiv -\frac{\gamma}{2\alpha}$, $P_2 \equiv \frac{\beta}{2\alpha}$, $E_1 \equiv -(\gamma'-\frac{\beta\gamma}{2\alpha})$, and $E_2 \equiv (\alpha'-\frac{\beta^2}{4\alpha})$.
It is then inserted to Eq.~(\ref{eq:landau_field2}).
\begin{equation}
\label{eq:landau_field3}
F(P^*,Q^*_{\Gamma_4^-},\mathcal{E}') \equiv \mathcal{A}(Q_{X_1^-}Q_{X_4^+})^2 +\mathcal{B}Q_{X_1^-}Q_{X_4^+}\mathcal{E}' +\mathcal{C}\mathcal{E}'^2,
\end{equation}
where
\begin{equation}
\begin{split}
&\mathcal{A} = -\alpha P_1^2 - \frac{E_1^2}{4E_2} = -1.86\ \text{eV} \\
&\mathcal{B} = -P_1 + \frac{E_1P_2}{2E_2} = -0.21\ \mu C/\text{cm}^2 \\
&\mathcal{C} = -\frac{P_2^2}{4E_2}-\frac{1}{4\alpha} = -0.49\ [\mu C/\text{cm}^2]^2/\text{eV}.
\end{split}
\end{equation}
It implies that non-polar modes $Q_{X_1^-}$ and $Q_{X_4^+}$ are also indirectly coupled with the external field, and with each other by the HIFE mechanism. 

To determine the switching field strength, elastic energy contributions from the non-polar modes [Eq.~(\ref{eq:non-polar})] are considered. 
\begin{equation}
\begin{split}
F_{\text{non-polar}}(Q_{X_1^-}, Q_{X_4^+}) = \eta Q_{X_1^-}^2 + \eta' Q_{X_4^+}^2 + \lambda Q_{X_1^-}^4 +\lambda' Q_{X_4^+}^4 +\xi Q_{X_1^-}^2Q_{X_4^+}^2, 
\end{split}
\end{equation}
where $\eta,\eta' < 0$ and $\lambda,\lambda' > 0$. The $\xi$ represents an elastic coupling between  $Q_{X_1^-}$ and $Q_{X_4^+}$.
When $P^*$ and $Q^*_{\Gamma_4^-}$ are determined by $Q_{X_1^-}$ and $Q_{X_4^+}$, total free energy is
\begin{equation}
\begin{split}
F(Q_{X_1^-},Q_{X_4^+},\mathcal{E}') 
= \eta Q_{X_1^-}^2 + \eta' Q_{X_4^+}^2 + \lambda Q_{X_1^-}^4 +\lambda' Q_{X_4^+}^4 
+(\xi+\mathcal{A}) Q_{X_1^-}^2Q_{X_4^+}^2 
+\mathcal{B}Q_{X_1^-}Q_{X_4^+}\mathcal{E}' +\mathcal{C}\mathcal{E}'^2 .
\end{split}
\end{equation}
Let's denote the total energy of the system with the structure given by $(Q_{\Gamma_4^-},Q_{X_1^-},Q_{X_4^+})$ as $E(Q_{\Gamma_4^-},Q_{X_1^-},Q_{X_4^+})$.
\begin{equation}
\begin{split}
\eta + \xi + \mathcal{A} &= -2E_3\\
\lambda &= E_3,
\end{split}
\end{equation}
where $E_3 \equiv E(0,0,1) - E(1,1,1) = 119.943874\ eV$. Note that when $Q_{X_1^-}=0$ or $Q_{X_4^+}=0$, $Q^*_{\Gamma_4^-}=0$.
\begin{equation}
\begin{split}
\eta' + \xi + \mathcal{A} &= -2E_4\\
\lambda' &= E_4,
\end{split}
\end{equation}
where $E_4 \equiv E(0,1,0) - E(1,1,1) = 0.362790 eV$.
\begin{equation}
\eta + \eta' + \lambda + \lambda' + \xi + \mathcal{A} = -E_5 \equiv -(E(0,0,0) - E(1,1,1)) =\ -121.125702 eV
\end{equation}
As a result,
\begin{equation}
\begin{split}
&\lambda = 120\ \text{eV} \\
&\lambda' = 0.36\ \text{eV} \\
&\eta = -241\ \text{eV} \\
&\eta' = -1.54\ \text{eV} \\
&\xi = 2.68\ \text{eV}
\end{split}
\end{equation}

One can see that the coefficients for $X_1^-$ mode ($\eta$ and $\lambda$) are much larger than others.
Therefore, it would be desirable to consider the FE switching of $Q_{X_4^+}$ mode, whereas we can assume the fixed value of $Q_{X_1^-}=1$ during the FE switching. Furthermore, it switches both the polarization and magnetic moment.
Free energy as a function of $Q_{X_4^+}$ is
\begin{equation}
F(Q_{X_4^+}) = (\eta' + \xi + \mathcal{A})Q_{X_4^+}^2 + \lambda'Q_{X_4^+}^4 + \mathcal{B}\mathcal{E}'Q_{X_4^+}
\end{equation}
The switching field strength can be obtained by the following.
\begin{equation}
\begin{split}
|\mathcal{B}|v|\mathcal{E}_c| = \frac{4}{3}|\eta' + \xi + \mathcal{A}|^{\frac{3}{2}}\frac{1}{\sqrt{6\lambda}} = \frac{8}{3\sqrt{3}}E_4\\
|\mathcal{E}_c| = \frac{8E_4}{3\sqrt{3}v|\mathcal{B}|} = 4.95\ \text{V/\AA}
\end{split}
\end{equation}

We can compare this with the value from a much simpler approach, which considers the polarization $P$ as a primary order parameter. In this case, free energy gain at the equilibrium value of the polarization $P_0 = 0.20\ \mu C/\text{cm}^2$ is $E_4$.
\begin{equation}
|\mathcal{E}^{\text{simple}}_c| = \frac{8E_4}{3\sqrt{3}vP_0} = 5.07\ \text{V/\AA}
\end{equation}
It is well compatible with the $|\mathcal{E}_c|$.
In ideal case, $E_1=2E_2$ to make $Q^*_{\Gamma_4^-}=1$ at the equilibrium (See Eq.~(\ref{eq:landau_G4})), \textit{i.e.}, $\mathcal{B} = -P_1+P_2 = -P_0$. Thus, two expressions are equivalent in ideal case.
Note that when $\eta$ and $\lambda$ are comparable to $\eta'$ and $\lambda'$, switching field estimation can differ from that of the simple approach.

\subsection{Revisit of Inorganic Perovskite Ca$_3$Mn$_2$O$_7$}

\begin{figure}[h]
\centering
\includegraphics[width=0.7\textwidth]{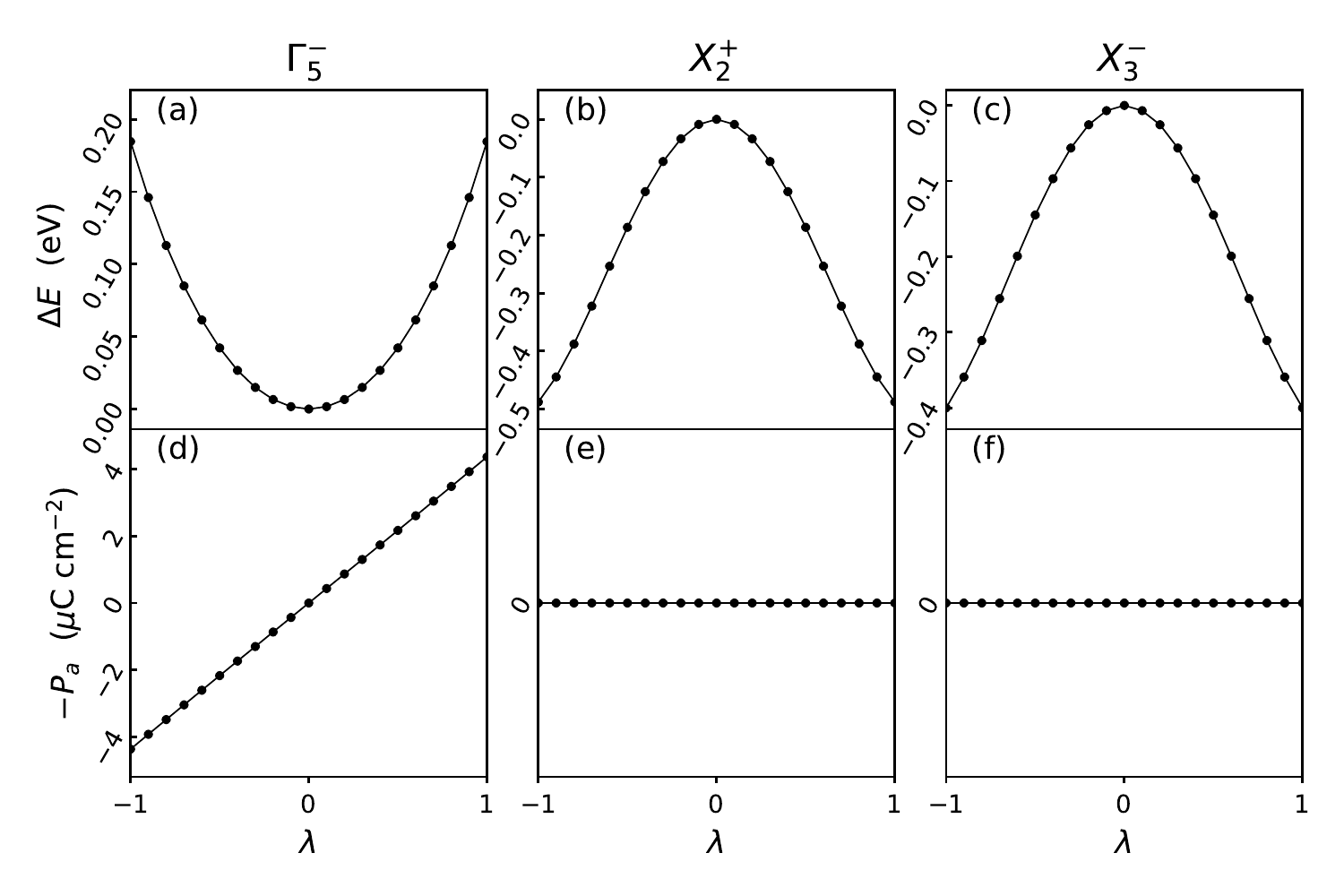}
\caption{
The case of Ca$_3$Mn$_2$O$_7$. (a-c) Change of energy and (d-f) electric polarization when only one of distortion mode among $\Gamma^-_5$, $X^+_2$ and $X^-_3$ exists.
}
\label{fig:CMO1}
\end{figure}

\begin{figure}[h]
\centering
\includegraphics[width=0.5\textwidth]{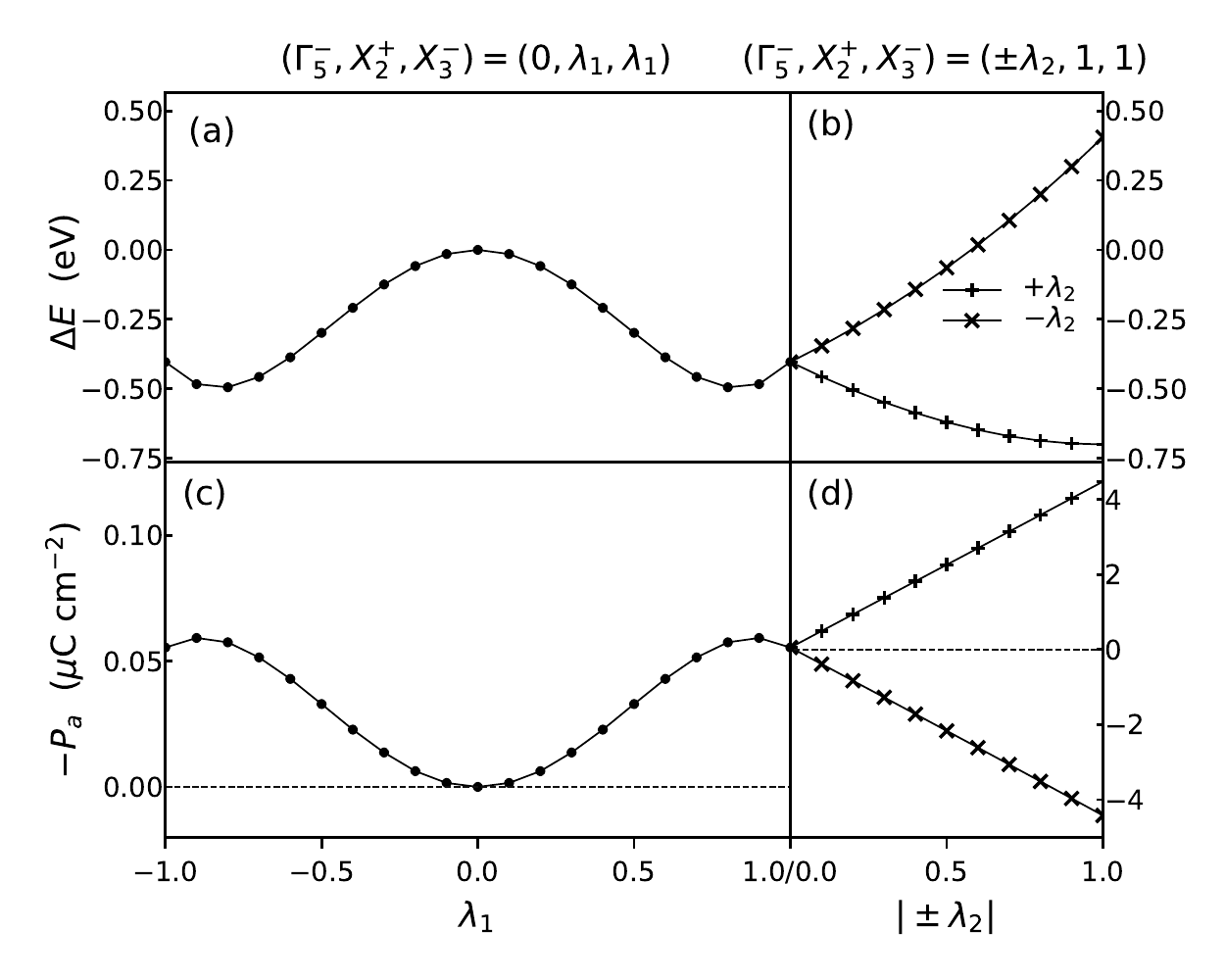}
\caption{
The case of Ca$_3$Mn$_2$O$_7$. (a,b) Change of energy, (c,d) electric polarization, and (e,f) magnetic moments along the parameter path $(\lambda_{\Gamma^-_5},\lambda_{X^+_2},\lambda_{X^-_3}) = (0, \lambda_1, \lambda_1)$ and $(\lambda_{\Gamma^-_5},\lambda_{X^+_2},\lambda_{X^-_3}) = (\pm\lambda_2, 1, 1)$. Be aware of the difference in the axis scale between (c) and (d).
}
\label{fig:CMO2}
\end{figure}

We revisited the prototypical inorganic perovskite HIFE material, Ruddlesden-Popper Ca$_3$Mn$_2$O$_7$~\cite{benedek_hybrid_2011}, in the view point of the free energy model we constructed in this work.
We adopt PBEsol+U functional and $4\times4\times4$ $k$-space grid within the primitive cell. Since we focus on the electric property, SOC is neglected. Other DFT parameters are the same as ref.~\cite{benedek_hybrid_2011}.
The polar structure and the corresponding non-polar structure have $A2_1am$ and $I4/mmm$ space group, respectively. These structures are connected by three distortion modes labeled by irreps $\Gamma^-_5$, $X^+_2$, and $X^-_3$.
$\Gamma^-_5$ mode is a polar mode that induces a polarization along $a$-direction. $X^+_2$ and $X^-_3$ modes are related to the rotation and tilt of the oxygen octahedra, respectively.
For the $A2_1am$ structure, experimental conventional cell lattice constants $a=5.2347$ \AA, $b=5.2421$ \AA, and $c=19.4177$ \AA\ are adopted~\cite{guiblin_ca3mn2o7_2002}. For the $I4/mmm$ structure, DFT optimized lattice constant $a=b=5.2320$ \AA\ and the same $c$ as $A2_1am$ structure are adopted. We assumed the linear interpolation of $a$ ($b$) with respect to the $\Gamma^-_5$ ($X^-_3$) mode.  

In Fig.~\ref{fig:CMO1} and \ref{fig:CMO2}, the total energy and polarization of Ca$_3$Mn$_2$O$_7$ with respect to the distortion modes are shown. In Fig.~\ref{fig:CMO1}, we can see the same behavior as Cu-MOF case, Fig.~\ref{fig:decomposed_1}. 
The polar mode $\Gamma^-_5$ is stable and induces a polarization. Two non-polar modes $X^+_2$ and $X^-_3$ are unstable and induce no polarization.
Fig.~\ref{fig:CMO2} which shows the effect of the hybrid mode $X^+_2\oplus X^-_3$ exhibits differences from Cu-MOF, Fig.~\ref{fig:decomposed_2}. The behavior of the total energy is similar. 
However, the hybrid mode induces only a small portion of the total polarization in comparison to the polar mode. In addition, the direction of the polarization by the hybrid mode and polar mode are the same. Thus, the switching of the polar mode inverts the sign of total polarization. 
One can say that the polarization of the Ca$_3$Mn$_2$O$_7$ behaves in an `intuitive' way.
These differences can be attributed to the difference between organic and inorganic nature.

The view point of the Landau theory provides a systematic comparison.
The parameters of the free energy described in the form of Eq.~(\ref{eq:landau1}) for the Ca$_3$Mn$_2$O$_7$ are determined in the same way except for $\alpha$. The $\alpha$ is determined by applying the external electric field to the crystal up to $\pm0.001$\ V/\AA\ via the method of Nunes and Gonze~\cite{nunes_berry-phase_2001}. Note that this method could not be applied for Cu-MOF due to a convergence issue. The obtained parameters are as follows.
\begin{equation}
\begin{split}
&\alpha = 1.62\times 10^{-3}\ (\text{eV}/[\mu C/\text{cm}^2]^2)\\
&\alpha' = 2.88\times 10^{-1}\ (\text{eV})\\
&\beta = 1.43\times 10^{-2}\ (\text{eV}/[\mu C/\text{cm}^2])\\
&\gamma = 1.80\times 10^{-4}\ (\text{eV}/[\mu C/\text{cm}^2])\\
&\gamma' = -5.52\times 10^{-1}\ (\text{eV}) .
\end{split}
\end{equation}
The $\beta$ and $\gamma$ represent the scales of the polarization by the polar mode and hybrid mode, respectively.
In the Cu-MOF case, $\beta\gamma<0$ and $|\beta|/|\gamma|\approx0.8<1$, whereas in the Ca$_3$Mn$_2$O$_7$, $\beta\gamma>0$ and $|\beta|/|\gamma|\approx80\gg1$.
When the sign of $\beta\gamma$ is positive (negative), the polarization by the hybrid mode and the polar mode are the same (opposite).
The ratio $|\beta|/|\gamma|$ tells us which mode contributes to the polarization larger. If this ratio is larger (smaller) than 1, the polar (hybrid) mode contribution is larger.

\subsection{Matrix Representations for $d$-orbitals Before and After The Jahn-Teller Transformation}
If we take $d$-orbitals $(l=2)$ as a basis in $\{d_{yz},d_{zx},d_{xy},d_{x^2-y^2},d_{z^2}\}$ order, matrix representations of $L_i$'s are as follows in the atomic units.

\begin{equation}
\begin{split}
L_{x}^{d} = 
\begin{pmatrix}
 0  & 0  & 0  & -i & -i\sqrt{3}  \\
 0  & 0  & i  & 0  & 0  \\
 0  & -i & 0  & 0  & 0  \\
 i  & 0  & 0  & 0  & 0  \\
 i\sqrt{3} & 0  & 0  & 0  & 0  \\
\end{pmatrix}
\end{split}
\end{equation}

\begin{equation}
\begin{split}
L_{y}^{d} = 
\begin{pmatrix}
 0  & 0  & -i & 0  & 0  \\
 0  & 0  & 0  & -i & i\sqrt{3}  \\
 i  & 0  & 0  & 0  & 0  \\
 0  & i  & 0  & 0  & 0  \\
 0  & -i\sqrt{3} & 0  & 0  & 0  \\
\end{pmatrix}
\end{split}
\end{equation}

\begin{equation}
\begin{split}
L_{z}^{d} = 
\begin{pmatrix}
 0  & i  & 0  & 0  & 0  \\
 -i & 0  & 0  & 0  & 0  \\
 0  & 0  & 0  & 2i & 0  \\
 0  & 0  & -2i & 0  & 0  \\
 0  & 0  & 0  & 0  & 0  \\
\end{pmatrix}
\end{split}
\end{equation}

The newly defined angular momentum operator matrices considering JT effect can be obtained by unitary rotation with this matrix, $(L_i)^{\text{new}} = U^{\dagger}(L_i)^{\text{old}}U$.

\begin{equation}
\label{eq:Lxdnew}
\begin{split}
(L_{x}^{d})^{\text{new}} = 
\begin{pmatrix}
 \quad0  & \quad0  & \quad0  &
i\sin(\tfrac{\theta_{\text{JT}}}{2})
-i\sqrt{3}\cos(\tfrac{\theta_{\text{JT}}}{2})&
-i\cos(\tfrac{\theta_{\text{JT}}}{2}) 
-i\sqrt{3}\sin(\tfrac{\theta_{\text{JT}}}{2})\\
 & \quad0  & \quad i  & 0  & 0  \\
 &  & \quad0  & 0  & 0  \\
 &  &  & 0  & 0  \\
\ \ h.c. &   &   &   & 0  \\
\end{pmatrix}
\end{split}
\end{equation}

\begin{equation}
\label{eq:Lydnew}
\begin{split}
(L_{y}^{d})^{\text{new}}=
\begin{pmatrix}
 \quad0  & \quad0  & \quad-i & 0  & 0  \\
   & \quad0  & \quad0  & 
i\sin(\tfrac{\theta_{\text{JT}}}{2}) 
+i\sqrt{3}\cos(\tfrac{\theta_{\text{JT}}}{2}) &  
-i\cos(\tfrac{\theta_{\text{JT}}}{2}) 
+i\sqrt{3}\sin(\tfrac{\theta_{\text{JT}}}{2})  \\
  &  & \quad0  & 0  & 0  \\
  &  &    & 0  & 0  \\
 \ \ h.c. &  &    &    & 0  \\
\end{pmatrix}
\end{split}
\end{equation}

\begin{equation}
\label{eq:Lzdnew}
\begin{split}
(L_{z}^{d})^{\text{new}} = 
\begin{pmatrix}
 \quad0  & \quad i  & \quad0  & 0  & 0  \\
   & \quad0  & \quad0  & 0  & 0  \\
   &    & \quad0  & 
-2i\sin(\tfrac{\theta_{\text{JT}}}{2}) & 
2i\cos(\tfrac{\theta_{\text{JT}}}{2})  \\
   &   &  & 0  & 0  \\
 \ \ h.c.  &   &  &   & 0  \\
\end{pmatrix}
\end{split}
\end{equation}

\subsection{Calculation of The Orbital Angular Momentum}
This appendix section describes the intermediate steps between Eq.~(\ref{eq:Li_dn}) and Eq.~(\ref{eq:Lxyz}).
If we sum up all the orbital angular momentum expectation values of the perturbed $d$-orbitals with up-spin, 
\begin{equation}
\sum_{n}\mel{d_{n\uparrow}}{L_i}{d_{n\uparrow}} 
= \sum_{n}\sum_{m\neq n}\Big[\tfrac{\mel{d^0_{m\uparrow}}{H_{SOC}}{d^0_{n\uparrow}}}{E^0_{n\uparrow}-E^0_{m\uparrow}}
\mel{d^0_{n\uparrow}}{L_i}{d^0_{m\uparrow}}+c.c.\Big] = 0
\end{equation}
because each of the terms is canceled with the term whose $n$ and $m$ are exchanged.
The same holds for down spin. It makes the calculation of orbital angular momentum for $d^4$ and $d^9$ easy. For $d^4$, $(L_i)_{d^4} = -\mel{d_{+\uparrow}}{L_i}{d_{+\uparrow}}$.
For $d^9$, from Eq.~(\ref{eq:H_SOC}) and $\mel{d^0_{n\uparrow}}{L_i}{d^0_{m\downarrow}} = 0$, $(L_i)_{d^9} = -\mel{d_{+\downarrow}}{L_i}{d_{+\downarrow}} = -(L_i)_{d^4}$.
The perturbed $\ket{d_{+}}$ is
\begin{equation}
\begin{split}
\ket{d_{+}} &= \ket{d^0_{+}}   
+ \left(\frac{\zeta}{2}\right)
\frac{\sin\theta\cos\phi
(-i\cos(\tfrac{\theta_{\text{JT}}}{2}) 
-i\sqrt{3}\sin(\tfrac{\theta_{\text{JT}}}{2}))}
{E^0_{+}-E^0_{yz}}\ket{d^0_{yz}} \\ 
&+ \left(\frac{\zeta}{2}\right)
\frac{\sin\theta\sin\phi
(-i\cos(\tfrac{\theta_{\text{JT}}}{2}) 
+i\sqrt{3}\sin(\tfrac{\theta_{\text{JT}}}{2}))}
{E^0_{+}-E^0_{zx}}\ket{d^0_{zx}} 
+ \left(\frac{\zeta}{2}\right)
\frac{2i\cos\theta
\cos(\tfrac{\theta_{\text{JT}}}{2})}
{E^0_{+}-E^0_{xy}}\ket{d^0_{xy}} .\\
\end{split}
\end{equation}
In the $d^4$ configuration,
\begin{equation}
\begin{split}
(L_i)_{d^4}&=-\mel{d_{+}}{L_i}{d_{+}}\\
&= -\mel{d_{+}^0}{L_i}{d_{+}^0} \\
&- \left(\frac{\zeta}{2}\right)
\frac{\sin\theta\cos\phi
(-i\cos(\tfrac{\theta_{\text{JT}}}{2}) 
-i\sqrt{3}\sin(\tfrac{\theta_{\text{JT}}}{2}))}
{E^0_{+}-E^0_{yz}} 
\mel{d_{+}^0}{L_i}{d^0_{yz}} + c.c. \\ 
&- \left(\frac{\zeta}{2}\right)
\frac{\sin\theta\sin\phi
(-i\cos(\tfrac{\theta_{\text{JT}}}{2}) 
+i\sqrt{3}\sin(\tfrac{\theta_{\text{JT}}}{2}))}
{E^0_{+}-E^0_{zx}} 
\mel{d_{+}^0}{L_i}{d^0_{zx}} + c.c.  \\
&- \left(\frac{\zeta}{2}\right)
\frac{2i\cos\theta
\cos(\tfrac{\theta_{\text{JT}}}{2})}
{E^0_{+}-E^0_{xy}}
\mel{d_{+}^0}{L_i}{d^0_{xy}} + c.c.  \\
\end{split}
\end{equation} 
Matrix elements $\mel{d_n^0}{L_i}{d_m^0}$ are given by Eq.~(\ref{eq:Lxdnew})-(\ref{eq:Lzdnew}).

\subsection{The Second Order Energy Correction Term in The Perturbation Theory }
In this appendix section, we calculate the energy correction by SOC in the perturbation approach which gives rise to the MSIA. 
In the previous study~\cite{stroppa_hybrid_2013}, only fixed JT phase and the same-spin contribution are considered for the MSIA. We improve the formulation by including the general JT phase and the opposite spin contribution. 
The first order energy correction $\mel{d^0_{n\sigma}}{H_{\text{SOC}}}{d^0_{n\sigma}}$ vanishes because the diagonal components of $L^d_i$ are 0. The lowest order energy correction is the second order correction term,
\begin{equation}
\Delta E^2_{\alpha} = \sum_{\beta\neq \alpha}\frac{|\mel{d^0_\beta}{H_{SOC}}{d^0_\alpha}|^2}{E^0_\alpha-E^0_\beta}. 
\end{equation}

It is convenient to separate the $H_{\text{SOC}}$ into the same-spin block and opposite-spin block.
\begin{equation}
\begin{split}
H_{\text{SOC},\uparrow\uparrow} 
&= \frac{\zeta}{2}(\sin\theta\cos\phi L_{x}
+  \sin\theta\sin\phi L_{y}
+  \cos\theta L_{x})
=-H_{\text{SOC},\downarrow\downarrow}
\end{split}
\end{equation}
and
\begin{equation}
\begin{split}
H_{\text{SOC,}\uparrow\downarrow} 
= \frac{\zeta}{2}\left[
(\cos\theta\cos\phi + i\sin\phi)L_{x} +
(\cos\theta\sin\phi - i\cos\phi)L_{y} +
(-\sin\theta)L_{z}
\right]
\end{split}
\end{equation}
The summation of the second order correction to the energy with the non-degenerate assumption in the half-filling case with only up spins is
\begin{equation}
\label{eq:2ndEcorrec1}
\begin{split}
\sum_{\alpha\in\uparrow}\Delta E^2_{\alpha} &= \sum_{\alpha\in\uparrow}\sum_{\beta\neq\alpha}
\frac{|\mel{d^0_{\beta}}{H_{SOC}}{d^0_{\alpha}}|^2}
{E^0_{\alpha}-E^0_{\beta}}\\
&= \sum_{n}\sum_{m\neq n}
\frac{|\mel{d^0_{m\uparrow}}{H_{SOC,\uparrow\uparrow}}{d^0_{n\uparrow}}|^2}
{E^0_{n\uparrow}-E^0_{m\uparrow}}
+\sum_{n}\sum_{m}
\frac{|\mel{d^0_{m\downarrow}}{H_{SOC,\downarrow\uparrow}}{d^0_{n\uparrow}}|^2}
{E^0_{n\uparrow}-E^0_{m\downarrow}}\\
\end{split} 
\end{equation}
The first term of the last line is the summation of the same-spin contribution to energy correction which vanishes. On the other hand, the second term is the summation of the opposite-spin contribution and is non-vanishing in general. It makes the difference between the expression of the energy correction of the $d^4$ spin configuration and that of the $d^9$. The opposite-spin contributions from each orbital of the $d^4$ configuration are as follows
\begin{equation}
\begin{split}
(\Delta E^2_{yz\uparrow})_{\downarrow\uparrow}
&=\left(\tfrac{\zeta}{2}\right)^2 [
\frac{\sin^2\theta}
{E^0_{yz\uparrow}-E^0_{zx\downarrow}}
+\frac{\cos^2\theta\sin^2\phi+\cos^2\phi}
{E^0_{yz\uparrow}-E^0_{xy\downarrow}} \\
&+\frac{(-\sin(\theta_{\text{JT}}/2)
+\sqrt{3}\cos(\theta_{\text{JT}}/2))^2
(\cos^2\theta\cos^2\phi+\sin^2\phi)}
{E^0_{yz\uparrow}-E^0_{-\downarrow}} \\
&+\frac{(\cos(\theta_{\text{JT}}/2)
+\sqrt{3}\sin(\theta_{\text{JT}}/2))^2
(\cos^2\theta\cos^2\phi+\sin^2\phi)}
{E^0_{yz\uparrow}-E^0_{+\downarrow}}
]
\end{split} 
\end{equation}

\begin{equation}
\begin{split}
(\Delta E^2_{zx\uparrow})_{\downarrow\uparrow}
&=\left(\tfrac{\zeta}{2}\right)^2 [
\frac{\sin^2\theta}
{E^0_{zx\uparrow}-E^0_{yz\downarrow}}
+\frac{\cos^2\theta\cos^2\phi+\sin^2\phi}
{E^0_{zx\uparrow}-E^0_{xy\downarrow}} \\
&+\frac{(\sin(\theta_{\text{JT}}/2)
+\sqrt{3}\cos(\theta_{\text{JT}}/2))^2
(\cos^2\theta\sin^2\phi+\cos^2\phi)}
{E^0_{zx\uparrow}-E^0_{-\downarrow}} \\
&+\frac{(\cos(\theta_{\text{JT}}/2)
-\sqrt{3}\sin(\theta_{\text{JT}}/2))^2
(\cos^2\theta\sin^2\phi+\cos^2\phi)}
{E^0_{zx\uparrow}-E^0_{+\downarrow}}
]
\end{split} 
\end{equation}

\begin{equation}
\begin{split}
(\Delta E^2_{xy\uparrow})_{\downarrow\uparrow}
&=\left(\tfrac{\zeta}{2}\right)^2 [
\frac{\cos^2\theta\sin^2\phi+\cos^2\phi}
{E^0_{xy\uparrow}-E^0_{yz\downarrow}} 
+\frac{\cos^2\theta\cos^2\phi+\sin^2\phi}
{E^0_{xy\uparrow}-E^0_{zx\downarrow}} \\
&+\frac{4\sin^2(\theta_{\text{JT}}/2)\sin^2\theta}
{E^0_{xy\uparrow}-E^0_{-\downarrow}} 
+\frac{4\cos^2(\theta_{\text{JT}}/2)\sin^2\theta}
{E^0_{xy\uparrow}-E^0_{+\downarrow}} 
]
\end{split} 
\end{equation}

\begin{equation}
\begin{split}
(\Delta E^2_{-\uparrow})_{\downarrow\uparrow}
&=\left(\tfrac{\zeta}{2}\right)^2 [
\frac{(-\sin(\theta_{\text{JT}}/2)
+\sqrt{3}\cos(\theta_{\text{JT}}/2))^2
(\cos^2\theta\cos^2\phi+\sin^2\phi)}
{E^0_{-\uparrow}-E^0_{yz\downarrow}} \\
&+\frac{(\sin(\theta_{\text{JT}}/2)
+\sqrt{3}\cos(\theta_{\text{JT}}/2))^2
(\cos^2\theta\sin^2\phi+\cos^2\phi)}
{E^0_{-\uparrow}-E^0_{zx\downarrow}} 
+\frac{4\sin^2(\theta_{\text{JT}}/2)\sin^2\theta}
{E^0_{-\uparrow}-E^0_{xy\downarrow}}
]
\end{split} 
\end{equation}
The spin direction dependent terms including $\theta$ and $\phi$ can be rewritten in physically intuitive expressions,
\begin{equation}
\begin{split}
&\cos^2\theta\cos^2\phi+\sin^2\phi
= 1-\sin^2\theta\cos^2\phi
= 1-(\hat{\mathbf{s}}\cdot\mathbf{x})^2 \\
&\cos^2\theta\sin^2\phi+\cos^2\phi
= 1-\sin^2\theta\sin^2\phi
= 1-(\hat{\mathbf{s}}\cdot\mathbf{y})^2  \\
&\sin^2\theta = 1-\cos^2\theta
= 1-(\hat{\mathbf{s}}\cdot\mathbf{z})^2,  \\
\end{split}
\end{equation}
where $\hat{\mathbf{s}}=\mathbf{s}/|\mathbf{s}|$ is the unit vector indicating the spin direction.
Meanwhile, it implies the existence of the spin direction independent contribution to the energy correction.
The same-spin contribution to the correction is simply $(\Delta E^2_{d^4})_{\uparrow\uparrow} = -(\Delta E^2_{+\uparrow})_{\uparrow\uparrow}$ because the first term of the Eq.~(\ref{eq:2ndEcorrec1}) vanishes.
\begin{equation}
\begin{split}
(\Delta E^2_{d^4})_{\uparrow\uparrow} &=
-\left(\tfrac{\zeta}{2}\right)^2
[
\frac{(\cos(\theta_{\text{JT}}/2)
+\sqrt{3}\sin(\theta_{\text{JT}}/2))^2
(\sin^2\theta\cos^2\phi)}
{E^0_{+\uparrow}-E^0_{yz\uparrow}} \\
&+\frac{(\cos(\theta_{\text{JT}}/2)
-\sqrt{3}\sin(\theta_{\text{JT}}/2))^2
(\sin^2\theta\sin^2\phi)}
{E^0_{+\uparrow}-E^0_{zx\uparrow}}
+\frac{4\cos^2(\theta_{\text{JT}}/2)\cos^2\theta}
{E^0_{+\uparrow}-E^0_{xy\uparrow}}
]\\
\end{split} 
\end{equation}

By summing up these terms, the spin direction dependent part of second order correction to the energy in $d^4$ configuration is
\begin{equation}
\label{eq:2ndEcorrectd4}
\begin{split}
\Delta E^2_{d^4}(\mathbf{S}) &=
\left(\tfrac{\zeta}{2}\right)^2
(\hat{\mathbf{s}}\cdot\mathbf{x})^2\Big[
\Big(\frac{-1}{E^0_{+\uparrow}-E^0_{yz\uparrow}}
+\frac{1}{E^0_{+\downarrow}-E^0_{yz\uparrow}}\Big)
(\cos(\theta_{\text{JT}}/2)
+\sqrt{3}\sin(\theta_{\text{JT}}/2))^2 \\
&+\Big(\frac{1}{E^0_{-\downarrow}-E^0_{yz\uparrow}}
+\frac{1}{E^0_{yz\downarrow}-E^0_{-\uparrow}}\Big)
(\sin(\theta_{\text{JT}}/2)
-\sqrt{3}\cos(\theta_{\text{JT}}/2))^2 %%\\
+\frac{1}{E^0_{xy\downarrow}-E^0_{zx\uparrow}}
+\frac{1}{E^0_{zx\downarrow}-E^0_{xy\uparrow}}
\Big]\\
&+\left(\tfrac{\zeta}{2}\right)^2
(\hat{\mathbf{s}}\cdot\mathbf{y})^2\Big[
\Big(\frac{-1}{E^0_{+\uparrow}-E^0_{zx\uparrow}}
+\frac{1}{E^0_{+\downarrow}-E^0_{zx\uparrow}}\Big)
(\cos(\theta_{\text{JT}}/2)
-\sqrt{3}\sin(\theta_{\text{JT}}/2))^2 \\
&+\Big(\frac{1}{E^0_{-\downarrow}-E^0_{zx\uparrow}}
+\frac{1}{E^0_{zx\downarrow}-E^0_{-\uparrow}}\Big)
(\sin(\theta_{\text{JT}}/2)
+\sqrt{3}\cos(\theta_{\text{JT}}/2))^2 %%\\
+\frac{1}{E^0_{xy\downarrow}-E^0_{yz\uparrow}}
+\frac{1}{E^0_{yz\downarrow}-E^0_{xy\uparrow}}
\Big]\\
&+\left(\tfrac{\zeta}{2}\right)^2
(\hat{\mathbf{s}}\cdot\mathbf{z})^2\Big[
\Big(\frac{-1}{E^0_{+\uparrow}-E^0_{xy\uparrow}}
+\frac{1}{E^0_{+\downarrow}-E^0_{xy\uparrow}}\Big)
4\cos^2(\theta_{\text{JT}}/2) \\
&+\Big(\frac{1}{E^0_{-\downarrow}-E^0_{xy\uparrow}}
+\frac{1}{E^0_{xy\downarrow}-E^0_{-\uparrow}}\Big)
4\sin^2(\theta_{\text{JT}}/2) %%\\
+\frac{1}{E^0_{zx\downarrow}-E^0_{yz\uparrow}}
+\frac{1}{E^0_{yz\downarrow}-E^0_{zx\uparrow}}
\Big]
\end{split} 
\end{equation}
For the $d^9$ spin configuration, the second order correction to the energy can be obtained simply.
\begin{equation}
\begin{split}
\Delta E^2_{d^9} &= 
\sum_{n=\text{occupied}}\Delta E^2_n = 
-\Delta E^2_{+\downarrow}\\
&= -\left(\tfrac{\zeta}{2}\right)^2
[
\frac{(\cos(\theta_{\text{JT}}/2)
+\sqrt{3}\sin(\theta_{\text{JT}}/2))^2
(\cos^2\theta\cos^2\phi+\sin^2\phi)}
{E^0_{+\downarrow}-E^0_{yz\uparrow}} \\
&+\frac{(\cos(\theta_{\text{JT}}/2)
-\sqrt{3}\sin(\theta_{\text{JT}}/2))^2
(\cos^2\theta\sin^2\phi+\cos^2\phi)}
{E^0_{+\downarrow}-E^0_{zx\uparrow}}
+\frac{4\cos^2(\theta_{\text{JT}}/2)\sin^2\theta}
{E^0_{+\downarrow}-E^0_{xy\uparrow}} \\
&+\frac{(\cos(\theta_{\text{JT}}/2)
+\sqrt{3}\sin(\theta_{\text{JT}}/2))^2
(\sin^2\theta\cos^2\phi)}
{E^0_{+\downarrow}-E^0_{yz\downarrow}} \\
&+\frac{(\cos(\theta_{\text{JT}}/2)
-\sqrt{3}\sin(\theta_{\text{JT}}/2))^2
(\sin^2\theta\sin^2\phi)}
{E^0_{+\downarrow}-E^0_{zx\downarrow}}
+\frac{4\cos^2(\theta_{\text{JT}}/2)\cos^2\theta}
{E^0_{+\downarrow}-E^0_{xy\downarrow}}
]\\
\end{split} 
\end{equation}
The spin direction dependent part is
\begin{equation}
\label{eq:2ndEcorrectd9}
\begin{split}
\Delta E^2_{d^9}(\mathbf{S}) &=
\left(\tfrac{\zeta}{2}\right)^2
(\hat{\mathbf{s}}\cdot\mathbf{x})^2\Big[
\Big(\frac{-1}{E^0_{+\downarrow}-E^0_{yz\downarrow}}
+\frac{1}{E^0_{+\downarrow}-E^0_{yz\uparrow}}\Big)
(\cos(\theta_{\text{JT}}/2)
+\sqrt{3}\sin(\theta_{\text{JT}}/2))^2 \Big]\\
&+\left(\tfrac{\zeta}{2}\right)^2
(\hat{\mathbf{s}}\cdot\mathbf{y})^2\Big[
\Big(\frac{-1}{E^0_{+\downarrow}-E^0_{zx\downarrow}}
+\frac{1}{E^0_{+\downarrow}-E^0_{zx\uparrow}}\Big)
(\cos(\theta_{\text{JT}}/2)
-\sqrt{3}\sin(\theta_{\text{JT}}/2))^2 \Big]\\
&+\left(\tfrac{\zeta}{2}\right)^2
(\hat{\mathbf{s}}\cdot\mathbf{z})^2\Big[
\Big(\frac{-1}{E^0_{+\downarrow}-E^0_{xy\downarrow}}
+\frac{1}{E^0_{+\downarrow}-E^0_{xy\uparrow}}\Big)
4\cos^2(\theta_{\text{JT}}/2)\Big]
\end{split} 
\end{equation}
Terms in the spin direction dependent part of the second order energy corrections are divided into three parts corresponding to the local coordinate directions. Each directional part is again divided according to the JT phase related factors.
Orbital ordering following the JT distortion affects the energy correction in two ways. One is the JT transformation of the $e_g$ orbitals which is explicitly expressed by the JT phase in Eq.~(\ref{eq:2ndEcorrectd4}) and Eq.~(\ref{eq:2ndEcorrectd9}). The other is the changes in the SOC-unperturbed orbital energies $E^0_i$ which is implicit in the expressions Eq.~(\ref{eq:2ndEcorrectd4}) and Eq.~(\ref{eq:2ndEcorrectd9}).
Because the crystal field splitting is larger than the changes by the JT effect, we can consider the factors in the trigonometric functions of the JT phase in the Eq.~(\ref{eq:2ndEcorrectd4}) and Eq.~(\ref{eq:2ndEcorrectd9}) are the leading factors to determine the MSIA direction. We can ignore the JT dependency of $E^0_i$'s for simplicity.
In $\Delta E^2_{d^4}(\mathbf{S})$, each directional parts are composed of three sub-parts. Two of them are JT dependent but they have different forms. The rest is JT phase independent. Because we assumed  $E^0_{e_{g}\uparrow}-E^0_{t_{2g}\uparrow}<E^0_{e_{g}\downarrow}-E^0_{t_{2g}\uparrow}$ and $E^0_{e_{g}\downarrow}-E^0_{t_{2g}\uparrow}>E^0_{t_{2g}\downarrow}-E^0_{e_{g}\uparrow}>0$, we can determine whether each of the JT phase dependent sub-part is the energy lowering or raising term according to the sign of these energy related factors. They are classified in Table.~\ref{tab:MSIA}. JT phase independent terms are always energy raising terms.
Energy lowering and raising terms result in the same tendencies of the favored spin direction.
On the other hand, $\Delta E^2_{d^9}(\mathbf{S})$ has only the energy lowering terms.

\begin{table}
\begin{tabular}{c|c|c}
\hline
 & $\Delta E < 0$
 & $\Delta E > 0$ \\
\hline
\hline
$(\hat{\mathbf{s}}\cdot\mathbf{x})^2$ &
$(\cos(\theta_{\text{JT}}/2)+\sqrt{3}\sin(\theta_{\text{JT}}/2))^2$ &
$(\sin(\theta_{\text{JT}}/2)-\sqrt{3}\cos(\theta_{\text{JT}}/2))^2$ \\
\hline
$(\hat{\mathbf{s}}\cdot\mathbf{y})^2$ &
$(\cos(\theta_{\text{JT}}/2)-\sqrt{3}\sin(\theta_{\text{JT}}/2))^2$ & 
$(\sin(\theta_{\text{JT}}/2)+\sqrt{3}\cos(\theta_{\text{JT}}/2))^2$ \\
\hline
$(\hat{\mathbf{s}}\cdot\mathbf{z})^2$ &
$4\cos^2(\theta_{\text{JT}}/2)$ &
$4\sin^2(\theta_{\text{JT}}/2)$ \\
\hline
\end{tabular}
\caption{
Classification of the JT phase dependent according to the energy lowering and raising}
\label{tab:MSIA}
\end{table}

\end{widetext}

\bibliography{CrCuMOF}

\end{document}